\documentclass[aip,reprint]{revtex4-1}
\usepackage{amsmath}
\usepackage{graphicx}        
\usepackage{amssymb}
\usepackage{bm}
\usepackage{siunitx}
\usepackage{standalone}
\usepackage{soul}
\usepackage{tikz}
\usepackage{expl3}
\usepackage{placeins}
\usetikzlibrary{calc,arrows,matrix}
\usepackage{xspace}
\usepackage{amsmath}           
\usepackage{amssymb}
\usepackage{pgfplots}
\usepackage{pgfplotstable}
\usepackage{booktabs}
\usepackage{float}
\usepgfplotslibrary{colorbrewer}
\usepgfplotslibrary{groupplots}
\usetikzlibrary{intersections}
\usepgfplotslibrary{fillbetween} 
\usetikzlibrary{shadows.blur}

\draft %

\begin{document}

\title[]{Direct stellarator coil optimization for nested magnetic surfaces with precise quasi-symmetry}
\author{Andrew Giuliani}
\email{giuliani@cims.nyu.edu}
\affiliation{ 
Courant Institute of Mathematical Sciences
New York University
New York, N.Y. 10012-1185
}%

\author{Florian Wechsung}%
\affiliation{ 
Courant Institute of Mathematical Sciences
New York University
New York, N.Y. 10012-1185
}%

\author{Antoine Cerfon}
\affiliation{ 
Courant Institute of Mathematical Sciences
New York University
New York, N.Y. 10012-1185
}%

\author{Matt Landreman}
\affiliation{ 
Institute for Research in Electronics and Applied Physics
University of Maryland
College Park, Maryland 20742, USA
}%

\author{Georg Stadler}
\affiliation{ 
Courant Institute of Mathematical Sciences
New York University
New York, N.Y. 10012-1185
}%

\date{\today}

\begin{abstract}
We present a robust optimization algorithm for the design of electromagnetic coils that generate vacuum magnetic fields with nested flux surfaces and precise quasi-symmetry.
The method is based on a bilevel optimization problem, where the outer coil optimization is constrained by a set of inner least-squares optimization problems whose solutions describe magnetic surfaces.  The outer optimization objective targets coils that generate a field with nested magnetic surfaces and good quasi-symmetry. The inner optimization problems identify magnetic surfaces when they exist, and approximate surfaces in the presence of magnetic islands or chaos. We show that this formulation can be used to heal islands and chaos, thus producing coils that result in magnetic fields with precise quasi-symmetry. 
We show that the method can be initialized with coils from the traditional two stage coil design process, as well as coils from a near axis expansion optimization.
We present a numerical example where island chains are healed and quasi-symmetry is optimized up to surfaces with aspect ratio 6.
Another numerical example illustrates that the aspect ratio of nested flux surfaces with optimized quasi-symmetry can be decreased from 6 to approximately 4. 
The last example shows that our approach is robust and a cold-start using coils from a near-axis expansion optimization.
\end{abstract}


\maketitle %

\section{Introduction}

Nested magnetic surfaces are not guaranteed to exist in stellarators, unlike in tokamaks, and confinement properties of the stellarator can be negatively affected by the presence of chaos and island chains in the magnetic field. Furthermore, the calculation of many physics performance metrics that might be used to optimize stellarators rely on the assumption of nested magnetic surfaces.
MHD stability, neoclassical confinement calculations, and optimization for quasi-symmetry on surfaces\cite{giuliani_wechsung_stadler_cerfon_landreman_2022} are simpler when the field presents nested magnetic surfaces. For all these reasons, it is crucial to develop stellarator optimization algorithms that are robust in the presence of chaotic field lines and island chains, and that will optimize the stellarator to a state with nested magnetic surfaces and precise quasi-symmetry, i.e., that can target island and chaos healing.  By ``precise" quasi-symmetry, we mean that the optimized coil designs generate magnetic fields that present an accurate approximation of quasi-symmetry where this property is targeted.

Quasi-symmetry is a favorable property of a magnetic field that ensures the collisionless trajectories of charged particles up to a certain energy will be confined \cite{helander_review}.
However, it is not completely known how closely this property can be approximated or realized in non-axisymmetric magnetic fields.
Recently, stellarators with very accurate (though imperfect) quasi-symmetry on a volume were discovered \cite{PhysRevLett.128.035001} using a novel stellarator optimization software framework \cite{simsopt}.

Island healing was a target during the design of the NCSX coils, where the aim was to find a plasma equilibrium and coil set such that specific resonant components of the field from the former and latter cancelled out \cite{Hudson_2002,Hudson_2003}.
In Ref.~\onlinecite{Zhu_2019}, a spectral analysis was conducted to manipulate island width size. An optimization-based approach was recently proposed, which relies on a combined VMEC-SPEC optimization of a toroidal boundary surface \cite{landreman2021stellarator}.
An initial configuration with islands and chaos was computed in a stage I optimization using VMEC.  Then, a combined VMEC-SPEC optimization was completed with the goal of optimizing away the significant island chains present in the initial state.  The VMEC solution was used to optimize for quasi-symmetry, and the SPEC solution was used to optimize away the magnetic islands by penalizing the island width via a Greene's residue computation.
The VMEC field was needed because the quasi-symmetry penalty term relies on the assumption of nested magnetic surfaces, which are not guaranteed in the SPEC field.
The latter field was needed because island chains cannot be resolved by the VMEC code.
It is also possible to only target island width using SPEC for finite $\beta$ equilibria, without optimizing for quasi-symmetry, which bore promising results\cite{baillod2022stellarator}.
One disadvantage of the Greene's residue approach is that one must know a priori which resonance one would like to heal.
The Greene's residue approach can be difficult to use if resonances enter or leave the magnetic configuration at a given iteration of the optimization, which happens particularly at the start of the optimization procedure.

Chaos healing also has been the focus of several studies in the past \cite{cary1982vacuum,dommaschk1982finite,cary1984construction,Hanson_1984_stochasticity,Cary1986_stochasticity,ejpauliota}. Instead of eliminating localized island chains corresponding to specific resonances, the idea here is to increase the plasma volume that is free from generalized stochasticity. This was first done for analytic vacuum fields, using perturbative analysis \cite{cary1982vacuum,dommaschk1982finite,cary1984construction}. Later, the Greene residue method was applied in an optimization framework to find coils generating vacuum fields with reduced chaos \cite{Hanson_1984_stochasticity,Cary1986_stochasticity}. These early works did not consider the simultaneous goal of having magnetic fields with a high level of quasi-symmetry. More recently, chaos healing was attempted in Ref.~\onlinecite{ejpauliota}, where a quadratic flux minimizing (QFM) penalty was added to a coil optimization objective to favor nested magnetic surfaces further away from the magnetic axis and optimize away chaotic regions of the field.  However, doing so does not control the quality of quasi-symmetry on low aspect ratio surfaces.

In this manuscript, we address some of the shortcomings of the previous approaches. Specifically, the main contributions of this work are: (1) we outline a robust approach for computing approximations of flux surfaces even in the presence of island chains and chaos, (2) based on that surface computation, we present a novel optimization-based approach to island and chaos healing, (3) our approach optimizes directly the geometry of electromagnetic coils, rather than a toroidal boundary surface, and (4) the algorithm promotes precise quasi-symmetry on nested magnetic surfaces.

In our previous work (Ref.~\onlinecite{giuliani_wechsung_stadler_cerfon_landreman_2022}), we outlined a numerical method to optimize stellarator coils for quasi-axisymmetry under the assumption that over the course of the optimization, the rotational transform remained strongly irrational.
Despite this strong requirement, we found precisely quasisymmetric magnetic fields generated by coils.
The procedure in Ref.~\onlinecite{giuliani_wechsung_stadler_cerfon_landreman_2022} relies on the solution to a partial differential equation (PDE) that can be difficult to solve numerically.
In this work, we propose a least-squares formulation to solve the PDE in a more robust fashion, resulting in a numerical optimization procedure that can be used even when nested flux surfaces do not exist.
Our approach is formulated as a bilevel optimization problem, where the outer coil optimization problem is constrained by the solution to a PDE, solved in a least squares sense in the inner optimization problem.
This is similar in spirit to the work in Ref.~\onlinecite{desc}, where quasisymmetric stellarators were found by solving an outer optimization problem constrained by the least squares solution to the force balance equations computed using the DESC code.
We also focus on curl-free magnetic fields as they are an important first step in the development of stellarator optimization algorithms.  Furthermore, vacuum fields can serve as suitable initial states in stellarator optimization approaches in which the plasma pressure is gradually ramped up\cite{boozer2019curl}.

\section{Computing surfaces}\label{sec:computingsurfaces}
The goal of this section is to outline a robust numerical method for computing magnetic surfaces in curl-free magnetic fields $\mathbf B \in \mathbb R^3$.
Even though the external magnetic fields that we use here are always
generated by electromagnetic coils, our method is not restricted to fields represented in this manner.
We use a finite-dimensional representation of a  
toroidal surface  $\bm \Sigma_s:[0,1)^2\rightarrow (x,y,z)$ that satisfies $n_\text{fp}$-rotational symmetry and stellarator symmetry.
$n_\text{fp}$ stands for the ``\textit{n}umber of \textit{f}ield \textit{p}eriods" and indicates the the number of times the field repeats itself after a full toroidal rotation.
The unknowns, or surface parameters,
in this representation are combined into a vector $\bm s\in \mathbb R^{n_s}$ where $n_s$ is the number of parameters that describe the surface; for details of this representation see Ref.~\onlinecite{giuliani_wechsung_stadler_cerfon_landreman_2022}.
Given a vacuum magnetic field $\mathbf{B}$, we seek to compute a magnetic surface in Boozer coordinates $\bm \Sigma_s(\mathbf{s})$, its rotational transform $\iota$, and the constant $G$, that solve $\mathbf{r}(\mathbf{s}) = [r_x(\mathbf s), r_y(\mathbf s), r_z(\mathbf s)]= 0$, where\cite{giuliani_wechsung_stadler_cerfon_landreman_2022} 
\begin{equation}\label{eq:pde}
    \mathbf{r}(\mathbf{s}) := G\frac{\mathbf{B}}{\|\mathbf{B}\|} - \|\mathbf{B}\| \left(\frac{\partial \bm \Sigma_s}{\partial \varphi} +\iota \frac{\partial \bm \Sigma_s}{\partial \theta}\right).
\end{equation}
This equation can be derived by equating two different representations
of the magnetic field, which assume that it is curl- and divergence-free \cite{giuliani_wechsung_stadler_cerfon_landreman_2022,boozer2019curl}.
Then, with the dual relations, it can be shown that magnetic surfaces parametrized in Boozer coordinates satisfy \eqref{eq:pde}.
Solutions to this partial differential equation can only be expected when $\iota$ is strongly irrational.
Based on this assumption, we have presented a pseudospectral approach to solve \eqref{eq:pde} in Ref.~\onlinecite{giuliani_wechsung_stadler_cerfon_landreman_2022}, and used that numerical method in an optimization loop to find coils with accurate quasi-symmetry.
The pseudospectral method aimed to find surfaces for which the residual was exactly zero at a fixed number of collocation points.
We called these surfaces ``BoozerExact" surfaces.
However, this numerical method can be brittle when nested flux surfaces do not exist, e.g. in regions with chaotic field lines and island chains, which occur at places where the rotational transform is not strongly irrational.
The approach that we describe now is more robust and can be used to determine surfaces even in regions where the pseudospectral method (BoozerExact) would have difficulty.

Discretizing this partial differential equation, surfaces are computed by solving the following constrained least squares minimization problem 
\begin{equation}\label{eq:surface-opt}
\begin{aligned}
    \min_{\mathbf{s}}~ & \frac 12 \int_{0}^1 \int_0^{1/n_{\text{fp}}} \|\mathbf{r}(\mathbf{s})\|^2~d\varphi ~d\theta\\
    \text{subject to } & V(\bm \Sigma_s) - V_0 = 0,
\end{aligned}
\end{equation}
where $\mathbf{r}(\mathbf{s}) :[0,1)\times[0,1/n_\text{fp})\to\mathbb R^3$,
\begin{equation}\label{eq:res}
\|\mathbf{r}(\mathbf{s})\|^2 = r_x(\mathbf{s})^2 +r_y(\mathbf{s})^2 + r_z(\mathbf{s})^2, 
\end{equation}
$V_0$ is a given target volume, and 
\begin{equation}\label{eq:vol}
V(\bm \Sigma_s) = \int_{0}^1\int_0^{1/n_{\text{fp}}}  \frac{1}{3} \bm{ \Sigma}_s \cdot \mathbf{n} ~d\varphi ~d\theta,
\end{equation}
where $\mathbf{n} = \partial \bm{ \Sigma}_s / \partial \varphi \times \partial \bm{ \Sigma}_s / \partial \theta$.  Note that while $\mathbf n$ is in the direction of the surface normal, here is not in general the unit normal.
The aim is to solve \eqref{eq:pde} in a least-squares sense by minimizing the quadratic residual.
The formula for the volume enclosed by the surface can be derived by recognizing
$$
V = \int_{D}~dx~dy~dz = \int_{D} \frac{1}{3} \nabla \cdot \vec{r} ~dx~dy~dz,
$$
where $\vec{r} = x \hat x + y \hat y + z \hat z \in D$ and $D$ is the region enclosed by the toroidal surface $\bm \Sigma_s$.  Applying the divergence theorem to the right-hand-side, we obtain \eqref{eq:vol} after substituting $\vec{r} = \bm \Sigma_s$.  

We use collocation on a tensor grid to approximate the integrals in \eqref{eq:surface-opt},
which results in the nonlinear least squares problem
\begin{equation}\label{eq:surface-opt-disc}
    \min_{\mathbf{s}}~ \frac{1}{6n_\varphi n_\theta} \|R(\mathbf{s})\|^2 + \frac{1}{2}w_v(V(\bm \Sigma_s) - V_0)^2
\end{equation}
where 
$$R(\mathbf{s}) = [r_x(\mathbf s)_1, r_y(\mathbf s)_1, r_z(\mathbf s)_1,\ldots, r_x(\mathbf s)_{N_\text{c}}, r_y(\mathbf s)_{N_\text{c}}, r_z(\mathbf s)_{N_{\text{c}}}],$$
$R(\mathbf s)\in \mathbb R^{3n_\varphi n_\theta}$ and the indices correspond to the collocation points.
We use a tensor product grid of $n_\varphi$ and $n_\theta$ collocation points in the $\varphi$ and $\theta$ directions, respectively.
We have experimented with two grids of quadrature points of the form $(\varphi_i, \theta_j) = (i\Delta \varphi, j\Delta \theta)$, for $i = 0, 1, \hdots, n_\varphi-1$, $j =0, 1, \hdots, n_\theta-1$. Rule 1 is on a full-period $(\varphi, \theta) \in [0, 1/n_{\text{fp}}) \times [0,1)$, $\Delta \varphi = (1/n_{\text{fp}})/n_\varphi$ and $\Delta \theta = 1/n_\theta$, $n_{\varphi} = 2n_{\text{tor}}+1$, $n_{\theta} = 2n_{\text{pol}}+1$.  This rule is spectrally accurate. Rule 2 is on a half-period $(\varphi, \theta) \in [0, 1/2 n_{\text{fp}}) \times [0,1)$ with $\Delta \varphi = (1/2n_{\text{fp}})/n_\varphi$, $\Delta \theta = 1/n_\theta$, $n_{\varphi} = n_{\text{tor}}+1$, $n_{\theta} = 2n_{\text{pol}}+1$.
This rule exploits stellarator symmetry, using half the number of points as rule 1 and is not spectrally accurate. A comparison of the two rules follows in Section \ref{sec:conv}.

A related concept is that of quadratic-flux-minimizing (QFM) surfaces
\cite{DEWAR199449}, which are surfaces that minimize
$$
\int_{0}^1\int_{0}^{1/n_{\text{fp}}} (\mathbf B \cdot \mathbf n)^2 ~d\theta d\varphi,
$$
without constraints on the angles that parametrize the surface.
QFM and BoozerLS surfaces are related to one another in regimes where nested flux surfaces exist.  In infinite dimensions, BoozerLS surfaces are also QFM surfaces, with the added requirement that the BoozerLS surface is parameterized in Boozer coordinates. 
In finite dimensions, numerical experiments show that BoozerLS surfaces do indeed approximate magnetic surfaces and we expect both BoozerLS and QFM to be close to each other.
The existence of QFM surfaces in more general contexts is delicate: when the quadratic flux is unweighted, it is shown in Ref.~\onlinecite{DEWAR199449} that only true flux surfaces extremize the QFM functional.
An analogous conclusion may be true for BoozerLS surfaces, but we have
not attempted to show this here.

The first-order optimality condition of \eqref{eq:surface-opt-disc} is 
\begin{equation}\label{eq:cons}
\mathbf{g}(\mathbf s):=J(\mathbf{s})^T R(\mathbf{s}) +
    w_v(V(\mathbf s) - V_0)\frac{\partial V}{\partial \mathbf s}  = \mathbf 0
\end{equation}
where $\mathbf g \in \mathbb R^{n_s}$, $J=\frac{\partial R}{\partial \mathbf s}\in \mathbb R^{(3n_\varphi n_\theta) \times n_{s}}$ is the Jacobian of $R(\cdot)$ and
\begin{equation}
    V(\mathbf s) := \frac{1}{3n_\varphi n_\theta}\sum_{i=1}^{3n_\varphi n_\theta} (\bm{ \Sigma}_s)_i \cdot \mathbf{n}_i.
\end{equation}
is the discretized surface volume.

In (Ref.~\onlinecite{giuliani_wechsung_stadler_cerfon_landreman_2022}), we showed numerically that computing BoozerLS surfaces is robust, even in the presence of islands and chaos. 
Framing the surface computation as a least-squares optimization problem allows us to use line search algorithms that track progress of the solution algorithm, and to prevent step sizes that are too large.  
Thanks to the optimization framework which defines the BoozerLS surfaces, we are free to introduce regularizations that prevent the numerically computed surfaces from self-intersecting.

The Newton step of \eqref{eq:cons} for the increment $\delta \mathbf s$ is
\begin{equation}\label{eq:Newton}
    \begin{aligned}
    \biggl[ J(\mathbf{s})^T J(\mathbf{s}) &+ w_v\frac{\partial V}{\partial \mathbf s}^T \frac{\partial V}{\partial \mathbf s} \\
    + \sum_{i=1}^{3n_\varphi n_\theta}R_i(\mathbf s)\frac{\partial^2 R_i}{\partial \mathbf s^2}(\mathbf s) &+  w_v(V(\mathbf s) - V_0)\frac{\partial^2 V}{\partial \mathbf s^2} \biggr] \delta \mathbf s  = -\mathbf g(\mathbf s)
    \end{aligned}
\end{equation}
where the Hessian matrix on the left-hand side multiplying $\delta \mathbf s \in \mathbb R^{n_s}$ is denoted by $H \in \mathbb R^{n_s \times n_s}$.
We initially use the Broyden–Fletcher–Goldfarb–Shanno (BFGS) algorithm to solve \eqref{eq:surface-opt-disc}.  Then, using this solution as an initial guess, we use Newton's method on \eqref{eq:Newton} to reduce the gradient of the nonlinear residual further. We observe that while the Hessian is not absolutely necessary for the computation of the least square surfaces, its availability is crucial for the computation of gradients in the outer coil optimization problem we consider in Section \ref{sec:coil-opt}, when the least-squares optimality condition is used as constraint within the coil optimization problem. We will highlight this point again in that section. We also note that in contrast to Ref. \onlinecite{giuliani_wechsung_stadler_cerfon_landreman_2022}, the residual in \eqref{eq:res} is divided by $\|\mathbf B \|$.  This is to prevent $\mathbf{r}$ from scaling with the coil currents.

The approach described in this section shows how to compute a magnetic surface that encloses a user-defined volume $V_0$.  The coil optimization studies in Section \ref{sec:coil-opt} use this numerical method to compute multiple magnetic surfaces, each with a different target volume.

\subsection{Convergence study}\label{sec:conv}
\begin{figure}
    \centering
    \includegraphics[width=\columnwidth]{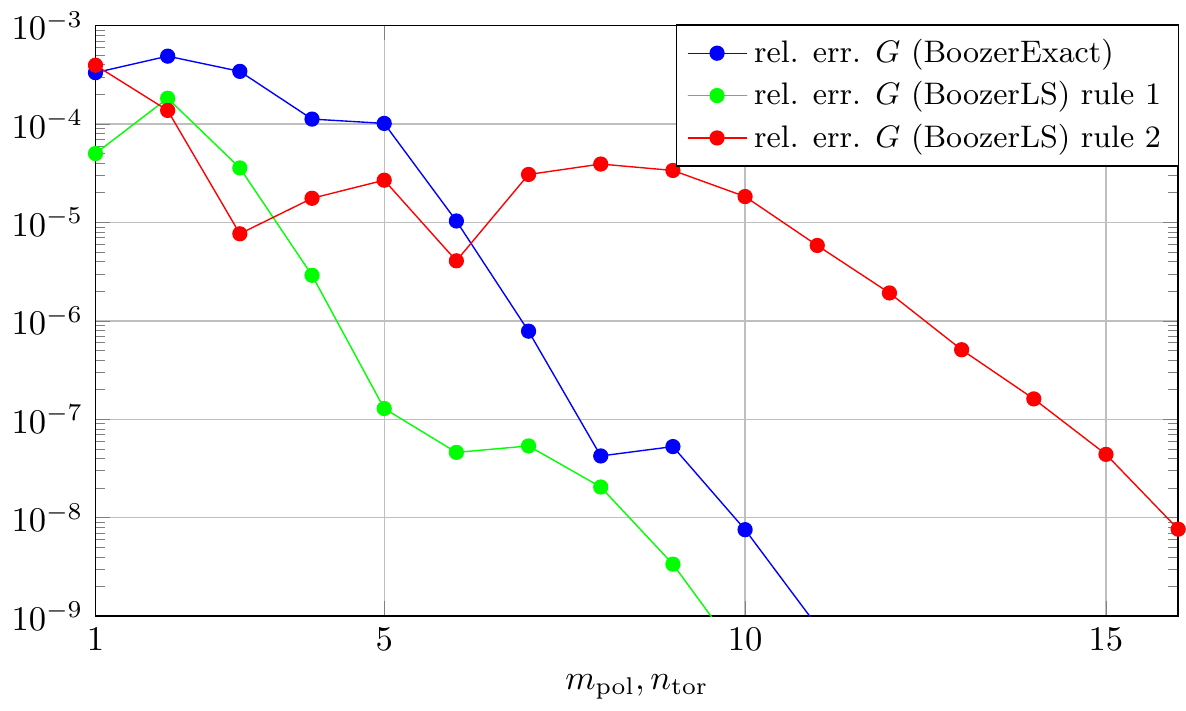}
    \caption{Convergence of $G$ with respect to number of surface modes on the innermost surface of the initial configuration in section \ref{sec:coilopt-1} and \ref{sec:coilopt-2}.
    Exponential convergence results in linear error curves on a $\log$-linear plot as observed here.
    Due to the spectral accuracy of the quadrature rule, we find that BoozerLS surfaces computed with rule 1 are more accurate than rule 2.  }
    \label{fig:error}
\end{figure}

In this section, we present a convergence test of the BoozerLS formulations using two different quadrature rules to approximate the integral in \eqref{eq:surface-opt}, called rules 1 and 2. 
We compare this new least squares formulation to the one in Ref.~\onlinecite{giuliani_wechsung_stadler_cerfon_landreman_2022}
i.e., when the number of unknowns coincides with the number of collocation points, which we refer to as BoozerExact.  We compute the innermost surface used in the coil optimization problem (section \ref{sec:coilopt}) using the BoozerExact and constrained BoozerLS formulations and vary the number of modes, $m_{\text{pol}}, n_{\text{tor}}$, in the surface representation and the number of collocation points.
Note that $m_{\text{pol}}$ and $n_{\text{tor}}$ are the number of Fourier modes used to represent the surface in the $\theta$ and $\varphi$ directions, respectively.
In addition to computing the unknowns that describe the geometry of the surface, we also compute its rotational transform $\iota$ as well as $G$.  For the configuration here, we know the exact value of $G = \mu_0 \sum_k I_k $, where $I_k$ is the current in the $k$th coil.
Thus, we can plot the convergence of the numerically computed $G$ to the true value, shown in figure \ref{fig:error}.
We find that the accuracy of the BoozerLS formulation depends on the choice of quadrature rule used to approximate the quadratic residual in \eqref{eq:cons}.
If a spectrally accurate quadrature rule is used (rule 1), then BoozerLS is as accurate as the BoozerExact formulation or better.
If a non-spectrally accurate quadratule is used (rule 2), then BoozerLS is orders of magnitude less accurate than the BoozerExact formulation.

Since the the BoozerLS formulation is most useful in regions without nested flux surfaces, where \eqref{eq:pde} does not even have well-defined solutions, we use rule 2 in what follows as it requires fewer points than rule 1.

\subsection{Surface regularization}
\begin{figure}
    \includegraphics[width=0.44\textwidth]{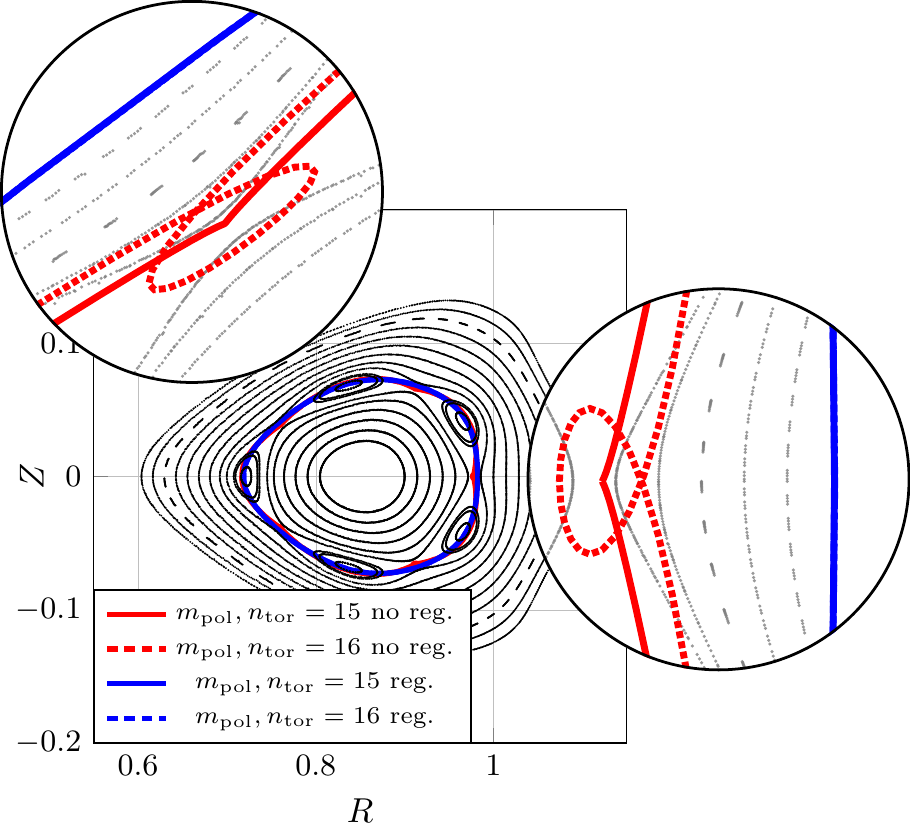}
    \caption{Cross sections for $m_{\text{pol}},n_{\text{tor}}=15, 16$ surfaces, plotted with solid and dashed lines, respectively.
    The red and blue cross sections correspond to the unregularized and regularized surfaces, respectively in the initial configuration of section \ref{sec:coilopt-1} and \ref{sec:coilopt-2}.
    The unregularized surface present a sharp cusp and self-intersections at the $X$-points, while the regularized surfaces do not.
    Moreover, the cross sections of the regularized surfaces overlap closely and cannot be distinguished visually.}\label{fig:xs_reg_noreg}
\end{figure}

\begin{figure}
    \includegraphics[width=0.44\textwidth]{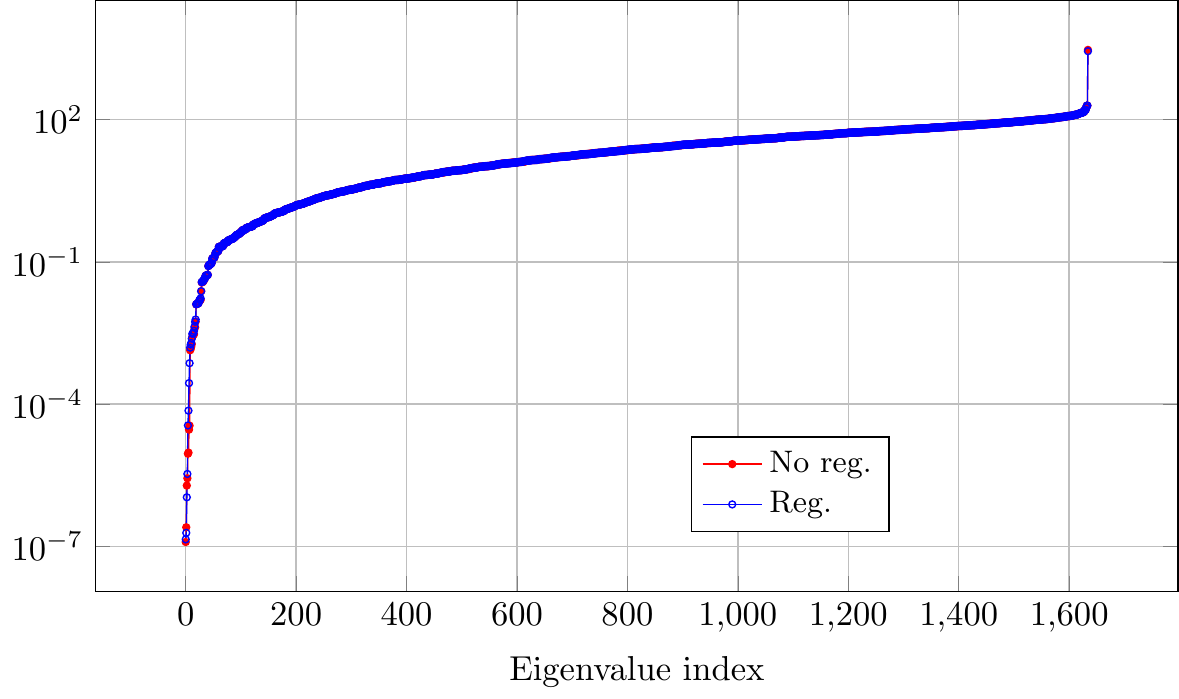}
    \caption{
    Spectrum of the Hessian at the minimizer of \eqref{eq:surface-opt-disc} when $m_{\text{pol}},n_{\text{tor}}=16$.  The are many small eigenvalues and as a result, there are many directions in which the surface can be perturbed without substantially increasing the magnitude of the residual $\| R(\mathbf s)\|^2$. 
    }\label{fig:spectrum}
\end{figure}

\begin{figure}
    \includegraphics[width=0.44\textwidth]{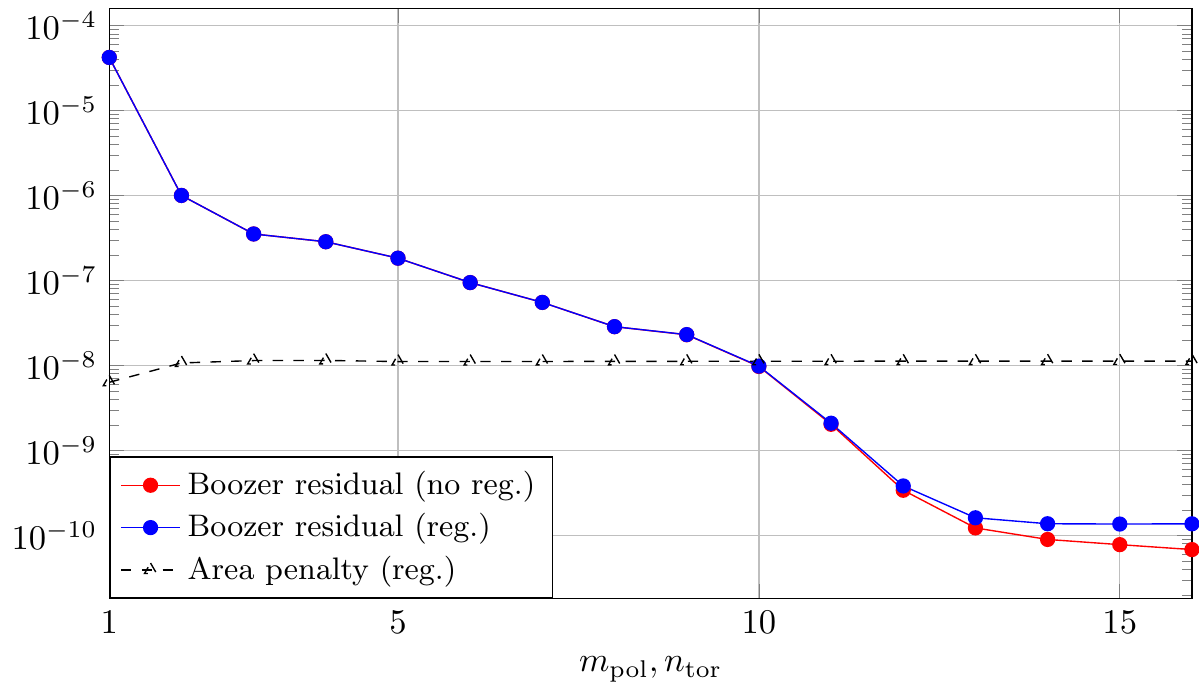}
    \caption{
    The convergence of the residual $\| R(\mathbf s)\|^2$ with increasing surface complexity $m_{\text{pol}}, n_{\text{tor}}$ with and without surface area regularization.
     As predicted by the eigenvalue analysis in Figure \ref{fig:spectrum}, the residual on the regularized surface still behaves like the unregularized one.
    }\label{fig:penalty}
\end{figure}

Numerical evidence suggests that surfaces determined by \eqref{eq:surface-opt} converge exponentially as a function of the number of Fourier modes that describe the surface in regions with nested flux surfaces.
However, we have not examined what happens in regions with chaos and islands.
As we increase the Fourier resolution, we observe that least squares surfaces may present self-intersections.
One possible surface regularization to prevent the formation of self-intersections is
$$
\frac{1}{2}w_v(A(\mathbf{s})-A_0)^2,
$$
where $A_0$ is 0.9 times the surface area on the unregularized surface, and $w_v>0$ is a weighting parameter for the regularization term.  
The goal of this regularization is to prevent the area of the surface from becoming too large for a given toroidal volume, thereby avoiding self-intersections.
The effect of this regularization term is illustrated in Figure \ref{fig:xs_reg_noreg}, where we compute a surface that passes through the $\iota=2/5$ island chain, for $m_{\text{pol}},n_{\text{tor}}=15,16$.
We observe that increasing the surface complexity from $m_{\text{pol}},n_{\text{tor}}=15$ to $m_{\text{pol}},n_{\text{tor}}=16$, results in surfaces that self-intersect in the neighborhood of the $X$-points.
We also compute the spectrum of the Hessian for the $m_{\text{pol}},n_{\text{tor}}=16$ surface (Figure \ref{fig:spectrum}).  
We find that before the regularization is added, there are many small eigenvalues and as a result, there is much freedom to add design requirements on the surface without affecting the magnitude of the Boozer residual.
This is confirmed by our numerical tests, as we observe that the regularization negligibly affects the Boozer residual penalty (Figure \ref{fig:penalty}).
The spectrum changes negligibly after adding the area penalty term, meaning that the door is open for additional regularizations to be included if needed.
The regularization introduced here is only a heuristic and there are no guarantees that self-intersections will always be prevented. We find that this solution is sufficient for our purposes, and useful at the start of a coil optimization procedure or if a BoozerLS surface is in the neighborhood of a low-order rational.
An alternative regularization is to penalize the surface curvatures, but we have not pursued this here for simplicity.

In what follows, the surface area regularization is used in section \ref{sec:coilopt-1} and \ref{sec:coilopt-2} to ensure robustness of coil optimization problems as the computed BoozerLS surfaces have a high number of Fourier modes ($m_{\text{pol}}, n_{\text{tor}}=15$).
Section \ref{sec:coilopt-3} does not require the surface area regularization for robustness, as BoozerLS surfaces with a lower number of Fourier modes ($m_{\text{pol}}, n_{\text{tor}}=10$) are computed.

\section{Coil optimization for quasi-symmetry on surfaces}

In this section, we show how to evaluate the quality of quasi-symmetry on surfaces, and we formulate a coil optimization problem that targets quasi-symmetry on those surfaces.  
Finally, we discuss the efficient computation of gradients for this optimization problem. 
In the previous section, we were concerned with developing a numerical method to approximate a single magnetic surface that encloses a user-provided volume.  In what follows, we use the numerical method to compute $N_s$ magnetic surfaces that enclose different volumes, and use those surfaces in a coil optimization procedure to control quasi-symmetry in the generated magnetic field.

\subsection{Measuring quasi-symmetry on surfaces}
Once a surface is known, we compute the deviation from quasi-symmetry on that surface following the approach from (Ref. \onlinecite{giuliani_wechsung_stadler_cerfon_landreman_2022}).  
In this work, we focus only on quasi-axisymmetry (QA), however, there is no fundamental restriction preventing the extension of this work to other types of quasi-symmetry.
First, the field strength $B(\varphi, \theta)$ on the surface is decomposed into a quasisymmetric and non-quasisymmetric component
$$
B(\varphi, \theta) = B_{\text{non-QA}}(\varphi, \theta) + B_{\text{QA}}(\theta).
$$
The quasisymmetric field strength is computed using a least squares projection
$$
B_{\text{QA}}(\theta) = \frac{\int^{1/n_{\text{fp}}}_{0} B(\bm\Sigma_s(\varphi,\theta))~\| 
\frac{\partial \bm\Sigma_s}{\partial \varphi} \times \frac{\partial \bm\Sigma_s}{\partial \theta} \|~d\varphi}{\int^{1/n_{\text{fp}}}_{0} ~ \| 
\frac{\partial \bm\Sigma_s}{\partial \varphi} \times \frac{\partial \bm\Sigma_s}{\partial \theta} \|~d\varphi},
$$
and is the closest quasisymmetric field strength to $B(\varphi, \theta)$ when measured in the norm $(\int_{\bm\Sigma_s} f^2 ~dS)^{1/2}$. The objective of the optimization problem presented in the next section uses this measure of non-quasi-symmetry to design stellarator coils. 

\subsection{The coil optimization problem}\label{sec:coil-opt}
Next, we formulate and solve an optimization problem for magnetic coils such that the induced magnetic surfaces have good quasi-symmetry. 
Where magnetic surfaces do not exist, our least squares framework will still provide a surface that can be used in the optimization procedure.
This is in contrast to our previous work\cite{giuliani_wechsung_stadler_cerfon_landreman_2022}, which would have had issues in regions in which the magnetic field does not possess nested flux surfaces.
At the end of the optimization algorithm, islands and generalized chaos will be healed.
We search for $N_c$ independent modular coils to which stellarator and rotational symmetries are applied such that the stellarator is made up of $2n_{\text{fp}}N_c$ modular coils. Each coil is parametrized with a Fourier representation and a current with $n_c$ degrees of freedom as in (Ref.~\onlinecite{giuliani2022single}). All coil degrees of freedom
are summarized into a vector $\mathbf c\in \mathbb R^{N_c n_c}$. We target quasi-symmetry of the induced magnetic field on $N_s$ surfaces, which are characterized as minimizers of \eqref{eq:surface-opt-disc} with the first-order necessary conditions
$$
\mathbf g_k(\mathbf s_k) = 0, k=1,\ldots,N_s
$$
as defined in \eqref{eq:cons}. 
Each surface satisfies the optimality condition, but has a different target volume.
In this way, we are able to extend the single-surface method described in Section \ref{sec:computingsurfaces} to multiple surfaces.
The corresponding surface is denoted by $\bm \Sigma_{s,k}$. Here and in the following, the index $k$ corresponds to the $k$th surface. All surface parameters are combined into a vector $\mathbf s:=(\mathbf s_1,\ldots,\mathbf s_{N_s} )$. The objective that we minimize is the sum of the average (normalized) non quasi-axisymmetry and the Boozer residuals on the surfaces $\bm \Sigma_{s,k}$:
\begin{equation}
\begin{aligned}
\hat{f}(\mathbf{c},\mathbf{s}) &:= \frac{1}{N_s}\sum^{N_s}_{k=1} \biggl\{\frac{\int_{\bm\Sigma_{s,k} }B_{\text{non-QA}}(\mathbf{c}, \mathbf{s}_k)^2~dS }{ \int_{\bm\Sigma_{s,k} }B_{\text{QA}}(\mathbf{c}, \mathbf{s}_k)^2~dS} \\
&+ \frac 12 w_r\int_{0}^1 \int_0^{1/n_{\text{fp}}} \|\mathbf{r}_k(\mathbf{s})\|^2~d\varphi ~d\theta \biggr\},
\end{aligned}\label{eq:objf4}
\end{equation}
where $w_r>0$ is a weighting parameter for the Boozer residual.  
If the Boozer residual is not in this objective, i.e. $w_r = 0$, then there is nothing preventing the accuracy of the least squares surface from degrading as the coils are optimized. 
Therefore, we include it to ensure that the least squares surface residual remains small, which is particularly important in the presence of islands and chaos, i.e., when no magnetic surfaces exist and the surface found with \eqref{eq:surface-opt-disc} is an approximate surface.
This term is not necessary for BoozerExact surfaces, when the same number of surface parameters and collocation points is used as in (Ref.~\onlinecite{giuliani_wechsung_stadler_cerfon_landreman_2022}).

This residual term can successfully detect regions of the magnetic field without nested flux surfaces\cite{giuliani_wechsung_stadler_cerfon_landreman_2022}.  When the residual term is small, then the BoozerLS surfaces accurately solve the PDE.  When it is large, then it is more likely that nested surfaces do not exist.  The addition of the residual term in the outer optimization problem will favor coils that produce nested magnetic surfaces, which solve \eqref{eq:pde} accurately.
This approach is similar to the one taken in Ref.~\onlinecite{ejpauliota}, where the quadratic flux minimizing (QFM) surface penalty was used to recover nested fluxed surfaces away from the magnetic axis without controlling for the quality of quasi-symmetry on those surfaces.
The technique that we adopt here differs in that not only can we favor nested flux surfaces with lower aspect ratio, but we can also directly optimize for quasi-symmetry on those surfaces.

Our goal is to find a set of coils that solves the following optimization problem:
\begin{subequations}\label{eq:optprob}
\begin{align}
    \min_{\mathbf c \in \mathbb R^{N_c n_{c}-1}, \: \mathbf s\in \mathbb R^{N_{\!s} n_s }} & ~ \hat f (\mathbf c, \mathbf s) \label{eq:optprob:J}\\
        \text{subject to }  \mathbf{g}_k(\mathbf{s}_k)=0 &\text{ for $k=1, \ldots, N_s$,}  \label{eq:optprob:2}\\
    c(\iota_1, \hdots, \iota_{N_s}) &= 0, \label{eq:optprob:3}\\
    R_{\text{major}} &= R_0, \label{eq:optprob:4}\\
    \sum_{i = 1}^{N_c} L^{(i)}_{c} &\leq L_{\max}, \label{eq:optprob:5}\\
    \kappa_i &\leq \kappa_{\max}, ~ i = 1, \ldots, N_c, \label{eq:optprob:6}\\
    \frac{1}{L^{(i)}_{c}}\int_{\bm\Gamma^{(i)}} \kappa_i^2 ~dl &\leq \kappa_{\mathrm{msc}}, ~ i = 1, \ldots, N_c, \label{eq:optprob:7}\\
    \| \bm\Gamma^{(i)}- \bm\Gamma^{(j)} \| &\geq d_{\min} ~ \text{ for } i \neq j,\label{eq:optprob:8} \\
    \|\bm \Gamma'^{(i)}\| - L^{(i)} &= 0 ~\text{ for } i = 1, \ldots, N_c, \label{eq:optprob:9}
\end{align}
\end{subequations}
where $c(\iota_1, \hdots, \iota_{N_s})$ is an equality constraint on the rotational transform profile that will be specified in the numerical examples.
The optimization is subject to the surface constraints \eqref{eq:optprob:2}--\eqref{eq:optprob:4} and the coil constraints \eqref{eq:optprob:5}--\eqref{eq:optprob:9}.
Note that in \eqref{eq:optprob:2}, we use the first-order optimality conditions \eqref{eq:cons} rather than the least squares minimization problem \eqref{eq:surface-opt-disc} to define surfaces. This is a standard approach to make bilevel optimization problems, i.e., optimization problems where the constraint is itself an optimization problem, computationally tractable.
Next, \eqref{eq:optprob:3} enforces that the rotational transform on a specific surface or that the average rotational transform is $\overline{\iota}$; other constraints on the rotational transform can easily be incorporated. The constraint \eqref{eq:optprob:4} fixes the major radius on the innermost surface to a given $R_0$ and prevents the length scale of the stellarator from changing.  \eqref{eq:optprob:5} prevents the sum of the independent modular coil lengths  $\sum_{i = 1}^{N_c} L^{(i)}_{c}(\mathbf{c})$ from exceeding a given value $L_{\max}>0$.  \eqref{eq:optprob:6} and \eqref{eq:optprob:7}, respectively, prevent the curvature and mean squared curvature on each coil from exceeding the values $\kappa_{\max}$ and $\kappa_{\mathrm{msc}}$.  \eqref{eq:optprob:8} ensures that the coils stay at least $d_{\min}>0$ away from one another and \eqref{eq:optprob:9} enforces that the coil parameterization has a uniform arclength.
Finally, in order to prevent the coil currents from approaching zero, the current in the first coil is fixed to a nonzero value over the course of the coil optimization, thus the dimensions of the outer and inner optimization problems are respectively $N_c n_c - 1$ and $N_s n_s$.

By virtue of the finite dimensional Fourier representation of the surfaces, we are restricted to magnetic surfaces that have some smoothness.
Thus, our optimized configurations will not likely present divertor surface shapes, even though it is known that they can be efficiently produced \cite{landreman2016efficient}.

\subsection{Handling of constraints and gradient computation}
The (in)equality constraints \eqref{eq:optprob:3}-\eqref{eq:optprob:8} are enforced by adding quadratic penalties to the objective. 
For example, to constrain the rotational transform on an inner surface to $2/5$, a quadratic penalty of the form $(\iota - 2/5)^2$ is added to the objective, where $\iota$ is the rotational transform on an interior surface.
In practice, the weights that multiply these quadratic penalties are only increased if the constraint is violated by more than 0.1\%.
Next, we discuss the computation of the discretely exact gradient of the reduced objective $f(\mathbf{c}) := \hat{f}(\mathbf{c}, \mathbf{s}(\mathbf{c}))$, i.e., the objective in which we consider the surfaces as a function of the coils through the magnetic field induced by the coils. We use an adjoint method to compute the gradient $\nabla_{\mathbf c} f$ efficiently as follows: 
\begin{subequations}\label{eq:grad-coil-optim}
\begin{align}
\nabla_{\mathbf c} f &= \frac{\partial \hat f}{\partial \mathbf{c}} -\sum_{k}^{N_s} \lambda_k^T \frac{\partial \mathbf g_k}{ \partial \mathbf{c}},\label{eq:grad-coil-optim:1}\\
H_k^T \lambda_k &= \frac{\partial \hat f}{\partial \mathbf{s}_k} ~\text{ for } k = 1, \hdots, N_s. \label{eq:grad-coil-optim:2}
\end{align}
\end{subequations}
Note that solving the $N_s$ adjoint systems \eqref{eq:grad-coil-optim:2}---one for each surface---requires the Hessian matrix from the Newton system \eqref{eq:Newton}. This is a consequence of the use of a least-squares formulation for the surface computation. Note that we typically finish the surface computation with a few Newton iterations, and thus the Hessian matrices $H_k$ needed in \eqref{eq:grad-coil-optim} are already available.

\subsection{Computational aspects} \label{sec:computational}
Many of the tools presented here are implemented in the SIMSOPT software package\cite{simsopt}, which is a suite for stellarator design and optimization.

The bilevel optimization problem described above can be computationally expensive. To illustrate this, consider first the inner optimization problems, which must be solved at each iteration of the outer optimization problem.
If $m_{\text{pol}},n_{\text{tor}}=N$, then there are $\mathcal{O}(N^2)$ surface degrees of freedom that are determined by the inner optimization problem.  Since $n_{\varphi} , n_{\theta} = \mathcal{O}(N)$, the number of residuals in the least squares objective scales like $\mathcal{O}(N^2)$.
As a result, the computational work to evaluate the gradient and Hessian of the Boozer residual in \eqref{eq:surface-opt} scales like $\mathcal{O}(N^4)$ and $\mathcal{O}(N^6)$, respectively.
These scalings are due to the use of a (globally defined) Fourier basis in the surface representation.
For example, increasing $N=10$ to $N=15$ will make the gradient and Hessian evaluation approximately 5 and 10 times more computationally expensive, respectively.
For these surface computations, we typically first use BFGS to robustly obtain a good approximate BoozerLS surface, and then improve the obtained solution using very few iterations of Newton's method.
As a result, the surface computation is largely dominated by the BFGS iterations.
At the expense of accuracy, some of the ill-scaling of the computational work could be mitigated by using more compact basis functions, e.g., finite element polynomial bases, which are generally not globally defined.
We also compared the performance of L-BFGS and BFGS algorithms for the surface computation, and, while each iteration of L-BFGS was faster, it typically required more iterations to reach the same accuracy as BFGS.

Consider now the outer coil optimization problem, which we solve using again the BFGS algorithm.
The computational work to evaluate the value of the objective and its gradient is dominated by the inner BFGS optimization and we observe that a single outer iteration takes under a minute when $N=10$ and over 5 minutes when $N=15$ on Intel Xeon Platinum 8268 processors, which agrees with the rough scaling computation in the previous paragraph.
This is in contrast to the local optimization algorithm presented in Ref.~\onlinecite{giuliani_wechsung_stadler_cerfon_landreman_2022}, where each iteration of the outer coil optimization algorithm took on the order of a few seconds when $N=10$.

Much effort has been made to accelerate the optimization algorithm.
MPI parallelism is used when quasi-symmetry is optimized on multiple surfaces, where each surface computation is completed on separate MPI ranks.
On each rank, the BoozerLS objective, gradient, and Hessian are evaluated using multiple cores with OpenMP in addition to SIMD parallelism on a given core.

\section{Experiments} \label{sec:coilopt}
In the following numerical experiments, we aim to demonstrate the robustness of the BoozerLS formulation and its ability to optimize for quasi-symmetry, even in the presence of islands and chaos.  The general optimization procedure we follow is to initially use BoozerLS surfaces everywhere.  Then, once the optimizer reaches the neighborhood of a local minimizer, surfaces in the neighborhood of low-order rationals continue to use the regularized BoozerLS formulation, while surfaces away from low-order rationals use the BoozerExact formulation.
The magnetic field that we design has two-fold rotational symmetry ($n_{\text{fp}} = 2$), stellarator symmetry, and a major radius $R_0=\SI{1}{\meter}$.  For all examples that follow, we set the coil design requirements to be $d_{\min}=\SI{0.1}{\meter}$, $\kappa_{\max} = \SI{5}{\meter}^{-1}$, $\kappa_{\text{msc}}=\SI{5}{\meter}^{-2}$, and $L_{\max} = \SI{18}{\meter}$.
We constrain the rotational transform profile to pass through the low-order rational $\iota=2/5$ and optimize for QA up to surfaces with aspect ratio 6 and 4 in sections \ref{sec:coilopt-1} and \ref{sec:coilopt-2}, respectively. 
In section \ref{sec:coilopt-3}, we constrain the average rotational transform in the volume confined by the outermost surface of aspect ratio 6 to be $0.42$.
Depending on the example, we optimize using eight or nine surfaces.  Using fewer surfaces might have resulted in comparable levels of quasi-symmetry\cite{giuliani_wechsung_stadler_cerfon_landreman_2022}, but we did not examine this question here.

In all examples, the $x, y, z$ coordinates of the independent modular coils are described using 16 Fourier modes resulting in $3(2\times 16+1) = 99$ geometric degrees of freedom and an associated current per coil, i.e. the number of degrees of freedom per coil is $n_c = 100$.
We design stellarators composed of four independent coils that have $4\times(99+1) = 400$ degrees of freedom.  
The outer coil optimization problem has $399$ dimensions however, since we fix the current on the first coil to be constant, thereby preventing the currents of the stellarator from approaching zero.
A complete set of 16 coils is obtained by applying twofold discrete rotational symmetry and stellarator symmetry.  For example \ref{sec:coilopt-1} and \ref{sec:coilopt-2}, we consider surfaces with $m_{\text{pol}},n_{\text{tor}}=15$, which results in a 1,439 dimensional inner optimization problem for each surface.
For example \ref{sec:coilopt-3}\color{black}, we consider surfaces with $m_{\text{pol}},n_{\text{tor}}=10$, which results in a smaller 661 dimensional optimization problem for each surface. 
See section \ref{sec:computational} for a detailed description of the design decisions made for the code to handle the complexity resulting from the bilevel optimization problem solved here.

The algorithm that we have followed in these experiments is:
\begin{enumerate}
    \item Obtain an initial coil set from a FOCUS-like\cite{zhu2018designing, wechsung2022precise} or near-axis expansion (NAE) optimization\cite{giuliani2022single}.
    \item Optimize the coils from step (1) for QA and nested flux surfaces using a number of BoozerLS surfaces.
    \item Optimize the coils from step (2) using the BoozerExact surfaces\cite{giuliani_wechsung_stadler_cerfon_landreman_2022} almost everywhere and keep BoozerLS surfaces only in the neighborhood of troublesome low-order rationals.
\end{enumerate}
As will be shown in the following examples, the magnetic field and physics properties of the stellarator change drastically during step (2).
For example, islands and chaotic regions may appear and disappear from one iteration to the next.
Despite this, the robust BoozerLS surfaces can still be computed and used for coil optimization.
At the end of step (2), we have reached a configuration with nested flux surfaces and much improved quasi-symmetry.
In step (3), we switch to the BoozerExact formulation as we are in the neighborhood of the optimizer and do not expect islands to appear anymore.
For the BoozerLS surfaces that remain, we still include the PDE residual penalty term in the objective, while for the BoozerExact surfaces, it is not needed.

We analyze the physics properties of the stellarator designs before and after optimization by examining plots of the non-QA ratio $\|B_{\text{non-QA}}\|/\|B_{\text{QA}}\|$, Boozer residual $\|\mathbf{r}\|^2_2$, rotational transform on surfaces in the toroidal volume.  
Each point on the profile corresponds to a BoozerLS surface that is computed with a continuation procedure.
The surfaces on which quasi-symmetry is optimized are verified by comparing their cross sections with Poincar\'e plots.
The physics profiles are generated by computing many BoozerLS surfaces through the volume and evaluating the physics quantities on each surface.
The physics plots are accurate away from island chains (Figure \ref{fig:error}) when computed using BoozerLS surfaces.
Even though the ratio $\|B_{\text{non-QA}}\|/\|B_{\text{QA}}\|$ can be computed, it is not well defined in regions of chaos and island as it relies on the assumption that nested flux surfaces exist in the underlying magnetic field.
Moreover, when the rotational transform is computed using BoozerLS surfaces, we find that the $\iota$ profile varies smoothly through island chains.  This is in contrast to what is observed when the profile is computed with field line tracing, where the rotational transform is constant through island chains\cite{landreman2021stellarator}.

\subsection{Island healing} \label{sec:coilopt-1}
\begin{figure}
    \centering
    \includegraphics[width=\linewidth]{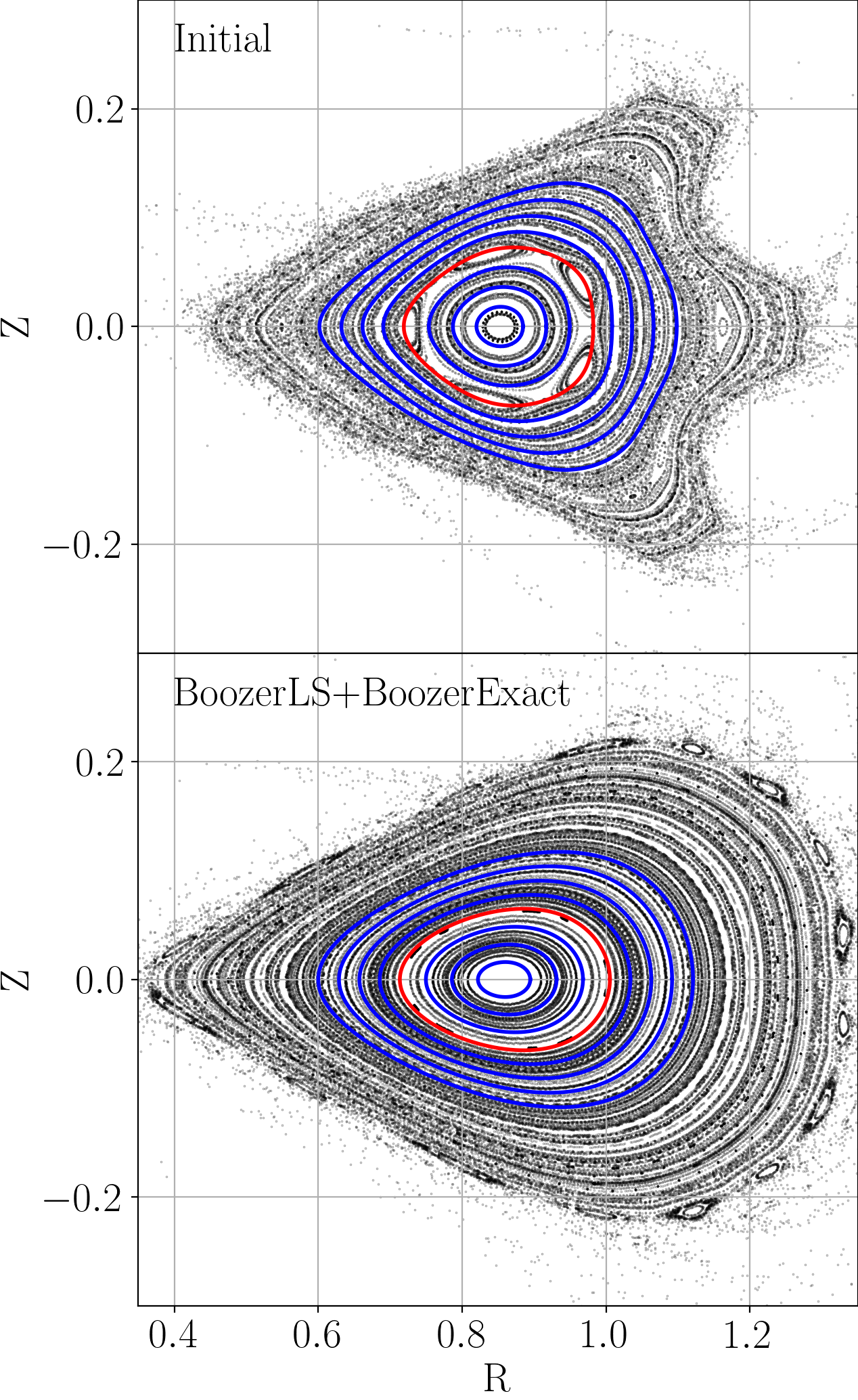}
    \caption{Island healing: Poincar\'e plot ($\phi=0$) and cross sections where we optimize for quasi-symmetry and nested flux surfaces.  
    The red cross section corresponds to the $\iota=2/5$ surface and the outermost blue surface has aspect ratio approximately 6.
    In the initial configuration, we use regularized BoozerLS surfaces for robustness.  After 1,000 iterations of the outer coil optimization, we switch to BoozerExact surfaces (blue) but keep the regularized BoozerLS surface at the low order rational $\iota=2/5$ (red).
    }
    \label{fig:physics_1a}
\end{figure}
\begin{figure}[h]
    \centering
    \includegraphics[width=\linewidth]{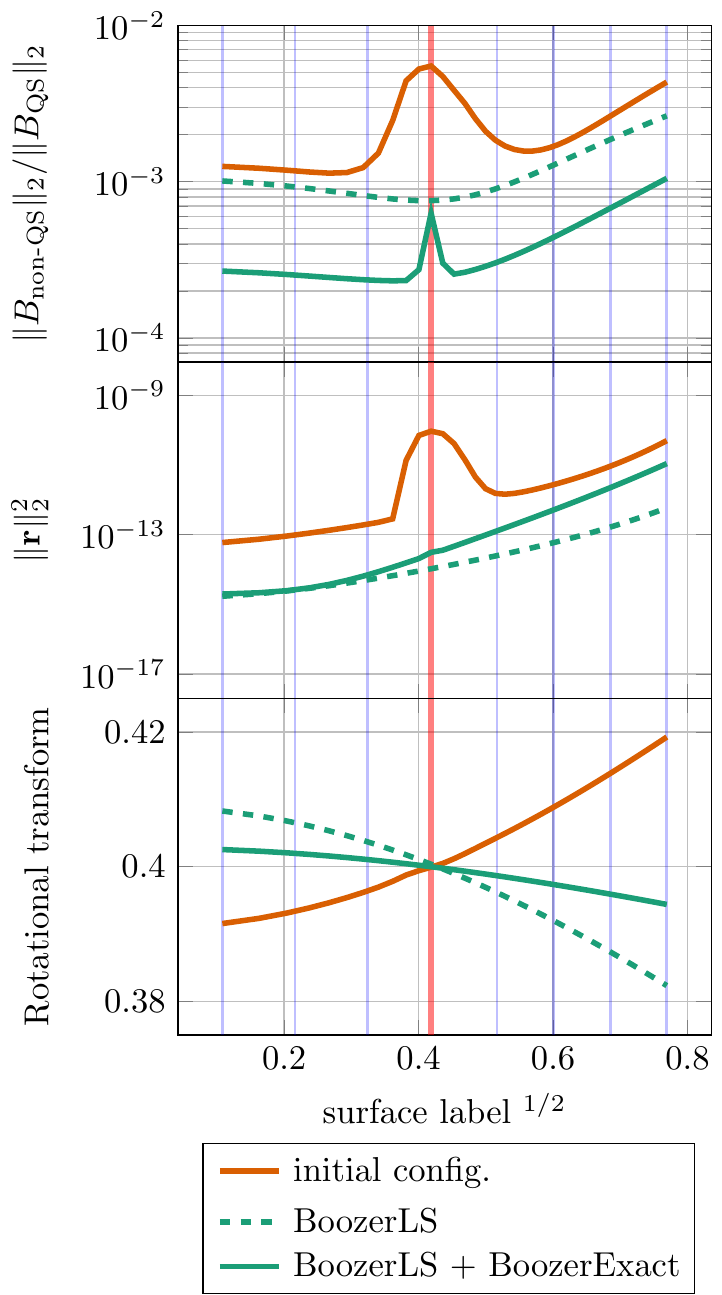}
    \caption{Island healing: physics quantities before and after island healing. The vertical red line corresponds to where the profile passes through $\iota=2/5$ and the vertical blue lines correspond to the other surfaces on which quasi-symmetry and nested flux surfaces were optimized.
    These figures were generated using least squares surfaces, with $m_{\text{pol}},n_{\text{tor}}=15$.
    We find that even though the island width has been greatly reduced, the extremum in the the non-QA ratio has not completely disappeared.
    The rotational transform profile varies smoothly through island chains when computed using BoozerLS surfaces, which is in contrast to what is observed when it is computed with field line tracing\cite{landreman2021stellarator}.
    }
    \label{fig:physics_1b}
\end{figure}
We begin with an initial equilibrium field 
presented in (Ref.\ \onlinecite{landreman2021stellarator}). This equilibrium was obtained by completing a stage I optimization for quasi-symmetry, with a target aspect ratio of 6.  Quasi-symmetry was optimized on a single surface at $0.5$ normalized toroidal flux, and the rotational transform was fixed to $0.39$ on the magnetic axis and to $0.42$ at the boundary.  That forces the rotational transform profile to pass through $2/5$, a low-order rational, resulting in an island chain.

We found a coil set for this equilibrium by solving a FOCUS-like stage II optimization problem (Refs.\ \onlinecite{zhu2018designing,wechsung2022precise}) to obtain an initial set of coils to reproduce the stage I equilibrium described in the previous paragraph.
The stage II optimization was launched 16 times, each with slightly perturbed initial guesses.
Once optimality was reached, the best performing coil set was chosen to initialize optimization described next.
Poincar\'e plots reveal a large island chain at $\iota=2/5$ (figure \ref{fig:physics_1a}).  
The non-QA ratio and BoozerLS residual through the plasma volume in this initial configuration are shown in Figure \ref{fig:physics_1b}, and we observe an extremum in both curves as the rotational transform passes through $\iota=2/5$.
This is expected as we are attempting to fit nested magnetic surfaces through a region with a significant island chain.
After 1,000 iterations of the outer coil optimization, the extremum is no longer visible and the slope of the rotational transform profile reverses.

The presence of the BoozerLS residual term in the objective results in a trade-off with the non-QA penalty term.
Switching to the faster BoozerExact formulation everywhere except the low-order rational surface allows the optimizer to further improve the quasi-symmetry in the volume. 
After the optimization, both the non-QA penalty and Boozer residual were further reduced.
We observe that the optimization successfully reduced the width of the island chain.
The rotational transform profile still passes through the low order rational number $2/5$ (figure \ref{fig:physics_1b}), and there still is a local extremum in the non-QA ratio plot, but its magnitude and width have been substantially reduced.

\subsection{Chaos healing}\label{sec:coilopt-2}
\begin{figure}
    \centering
     \includegraphics[width=\linewidth]{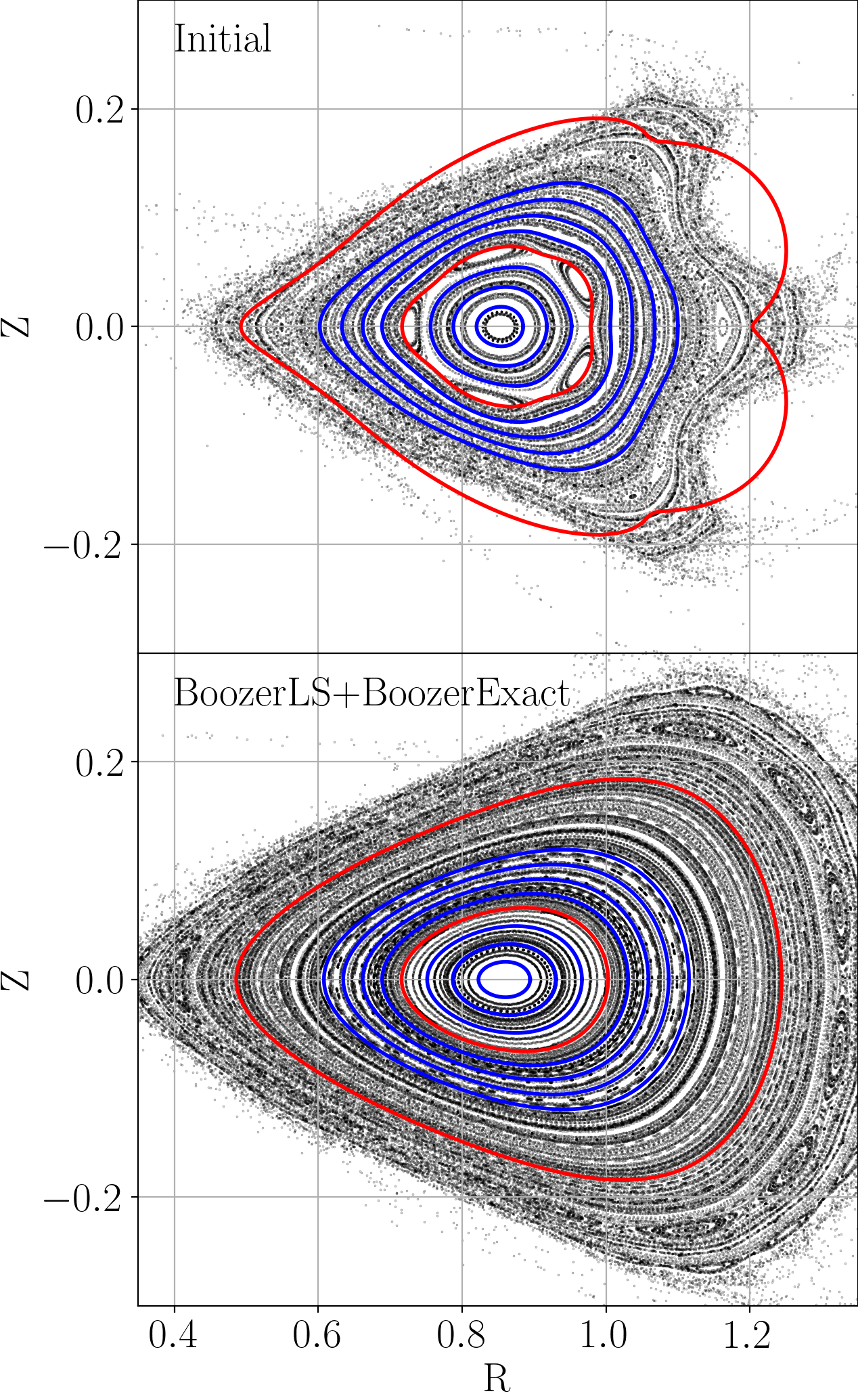}
    \caption{Chaos healing: Poincar\'e plot ($\phi=0$) and cross sections where we optimize for quasi-symmetry and nested flux surfaces.  
    The innermost and outermost red cross sections correspond respectively to the $\iota=2/5$ surface and the additional aspect ratio 4 surface. }
    \label{fig:physics_2a}
\end{figure}
\begin{figure}
    \centering
    \includegraphics[width=\linewidth]{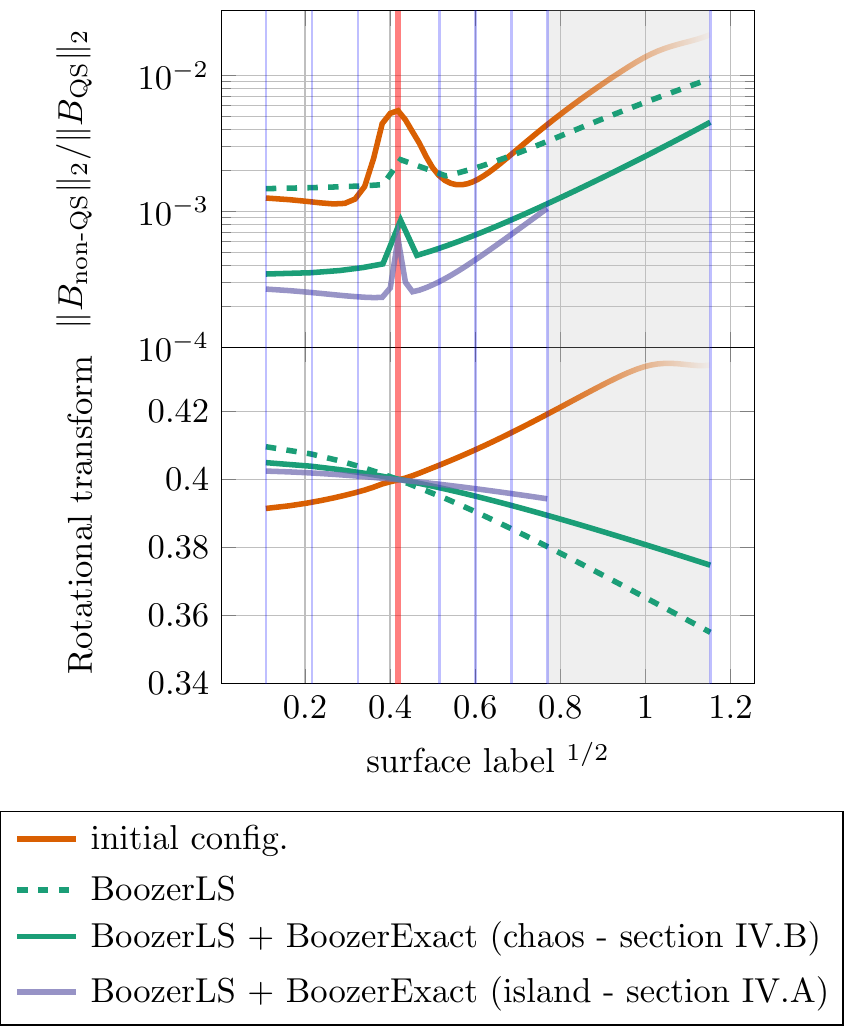}
    \caption{Chaos healing: physics quantities before and after optimization. The greyed region of the plot corresponds to the volume between the outermost blue and red cross section in Figure \ref{fig:physics_2a}.
    Since this is a region with chaotic field lines and without nested flux surfaces, the physics quantities are not well defined.
    The vertical red line corresponds to where the profile passes through $\iota=2/5$ and the vertical blue lines correspond to the surfaces on which we optimize for quasi-symmetry and nested flux surfaces.
    For comparison, we also include the results from the previous chaos healing example, where we only optimized on a smaller volume with aspect ratio 6.}
    \label{fig:physics_2b}
\end{figure}

The magnetic field generated by the initial coil set in section \ref{sec:coilopt-1} presents nested flux surfaces, a significant island chain, and chaotic field lines in lower aspect ratio regions.
The goal of this example is to simultaneously heal islands as well as increase the volume in which nested flux surfaces are present.
We do this by adding a ninth surface with aspect ratio approximately 4 to the optimization problem.

A Poincar\'e plot of the initial and optimized configurations, along with cross sections of the BoozerLS surfaces, is shown in Figure \ref{fig:physics_2a}. 
A discussion of the coils can be found in section \ref{sec:discussion}. We also stress the fact that in the initial configuration, the outermost surface passes through a chaotic regions of the magnetic field.  
We are able to compute such a BoozerLS surface thanks to the least squares framework adopted in this work.
Computing a BoozerExact surface through the same regions of the field would be difficult since clearly nested surfaces do not exist.

In Figure \ref{fig:physics_2b}, the physics properties of the initial and optimized stellarators are presented. Note that the non-QA ratio and rotational transform are not well defined for the outermost surface in the original configuration, which we indicate by the fading line. After initial coil optimization using the BoozerLS surface description, reasonable magnetic surfaces are available and thus we continue the coil optimization with the BoozerExact surface formulation. The resulting design has nested flux surfaces as can be seen visually from the Poincar\'e plot (bottom of Figure \ref{fig:physics_2a}), and reasonable non-QA ratio also out to the outermost surface (Figure \ref{fig:physics_2b}).
Comparing with the previous example where we optimized on a smaller volume, we find that requesting precise QA on lower aspect ratio surfaces causes the quality of QA to degrade slightly at the core.
This is unsurprising as the smaller the region on which quasi-symmetry is requested, the easier it is to find a magnetic field with that property \cite{landreman2021stellarator, giuliani2022single}.
In the next example, we present a configuration optimized for QA only on the magnetic axis, which only presents extremely good quasi-symmetry at the core, and whose quasi-symmetry quickly degrades moving away from the axis.

\subsection{Cold start direct coil optimization}\label{sec:coilopt-3}
In the previous two examples, we followed the classical two-stage optimization procedure to obtain coils, which we then improved using the optimization problem \eqref{eq:optprob}. In this final example we show that thanks to the BoozerLS formulation, this two stage procedure is unnecessary and we can design stellarators with nested surfaces and precise QA properties starting from flat coils. 

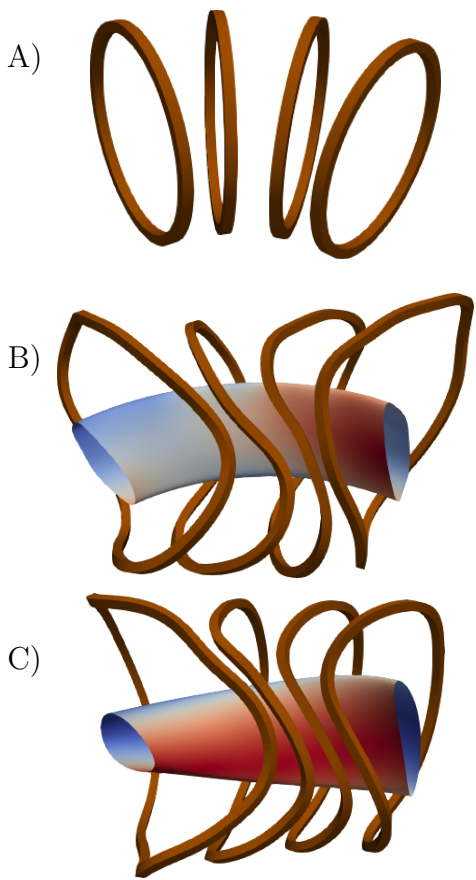
\begin{figure}
    \begin{tikzpicture}
    \node (A) at (-4,5) {\large A)};
    \node (B) at (-4,1) {\large B)};
    \node (C) at (-4,-3) {\large C)};    
    \node (Apic) at (-.8,4) {
    \includegraphics[height=0.14\textheight]{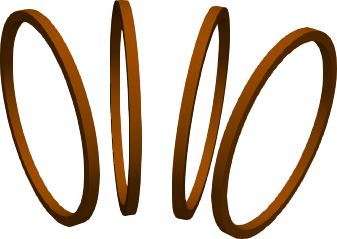}};
    \node (Bpic) at (-.8,0) {
    \includegraphics[height=0.16\textheight]{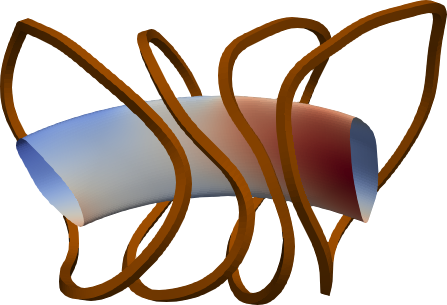}};
    \node (Cpic) at (-.8,-4) {
    \includegraphics[height=0.16\textheight]{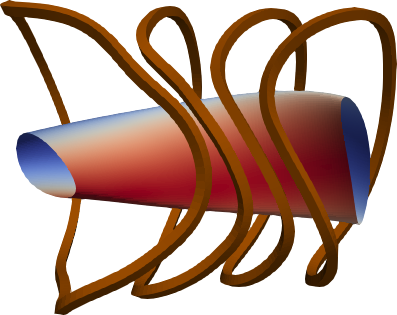}};
    \end{tikzpicture}
    \caption{Cold start direct coil optimization: Shown in (A) are the 
    flat coils with zero current used as initialization. The coils obtained with the near-axis expansion optimization are shown in (B). These coils are used as initialization for the BoozerLS and BoozerExact optimization, which result in the coils shown in (C). The optimized coils from a previous work (Ref.~\onlinecite{giuliani_wechsung_stadler_cerfon_landreman_2022}) obtained with a different local optimization method are visually indistinguishable from the coils in (C).  The coil currents of the two designs are remarkably close to each other as well (Table \ref{tab:currents}).
    }
    \label{fig:nae_coils}
\end{figure}

The computation of Boozer surfaces in a magnetic field is more straightforward when a magnetic axis with some nested flux surfaces are present.
Moreover, the solution to \eqref{eq:pde} is not unique when $\iota=0$. To avoid these issues, we start with an initial configuration obtained from a near axis expansion (NAE) optimization\cite{giuliani2022single} beginning from equispaced, flat coils with zero current (Figure \ref{fig:nae_coils}). 

In this initial NAE optimization, the geometry of the magnetic axis and electromagnetic coils are optimized for QA on the magnetic axis. The coil design requirements we impose are similar to those in Section \ref{sec:coilopt-1}, except that we constrain each coil to have the same length of $L=\SI{4.5}{\meter}$ rather than constraining the total coil length to be less than $L_{\max}=\SI{18}{\meter}$. We found that when doing the latter, the coils tend to have lengths that are disparate from one another as only physics properties on the magnetic axis (as opposed to in the volume) are targeted.
The rotational transform on axis is constrained to $\iota=0.42$ and  the mean radial position of the axis is constrained to be 1.

\begin{figure}
    \centering
    \includegraphics[width=\linewidth]{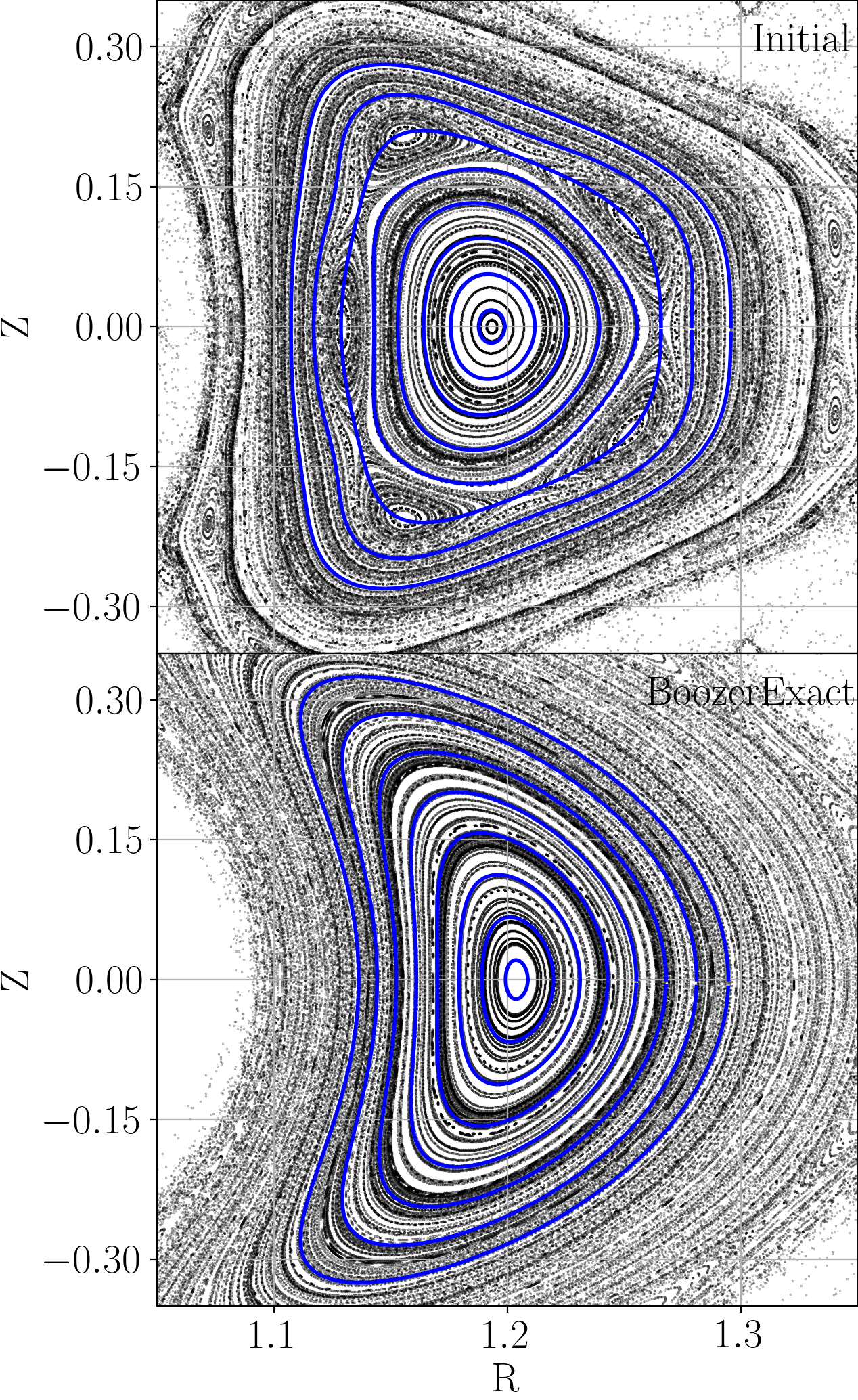}
    \caption{Cold start direct coil optimization: Poincar\'e plot ($\phi=0$) in the configuration obtained from the near axis expansion and after BoozerExact optimization.
    The cross sections (blue curves) in the initial and final configurations correspond to the surfaces used in the BoozerLS and BoozerExact phases of the optimization.
    Despite the presence of island chains in the initial configuration, the BoozerLS optimization successfully replaces them with nested flux surfaces with precise quasi-symmetry.
    }
    \label{fig:physics_3a}
\end{figure}
\begin{figure}
    \centering
    \includegraphics[width=\linewidth]{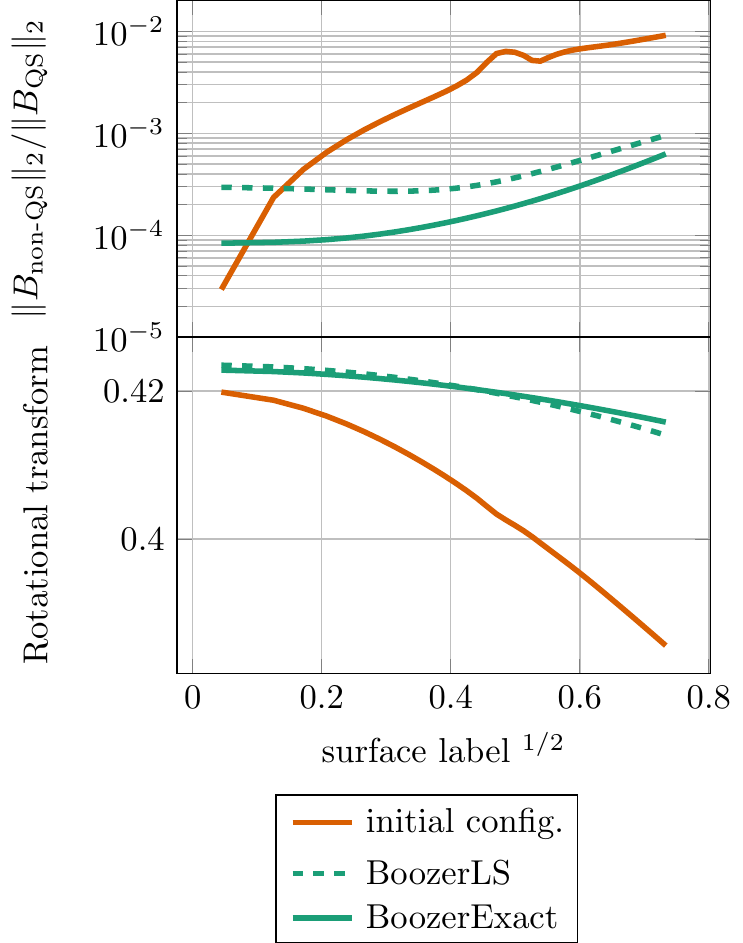}
    \caption{Cold start direct coil optimization: physics quantities after optimization from cold start.}
    \label{fig:physics_3b}
\end{figure}

The NAE optimization is robust, and from a cold start finds a configuration with precise quasi-symmetry on axis.
However, the quality of the quasi-symmetry quickly degrades as one moves away from the axis (Figure \ref{fig:physics_3b}), as also observed in Ref.\ \onlinecite{giuliani2022single}.  
The rotational transform on axis in the optimized configuration decreases from 0.42 and crosses the low order rational $\iota=2/5$, which results in an island chain (Figure \ref{fig:physics_3a}).

Using the coil set from the NAE optimization, we use eight regularized BoozerLS surfaces, which pass directly through the island chain.  
After 1,000 iterations of BFGS, the islands have been healed and the rotational transform no longer crosses a low order rational. Thus, we no longer need to use BoozerLS surfaces and switch all the surfaces to BoozerExact.
Since now there no longer is a trade-off between the non-QA penalty and the Boozer residual in the objective, one can more effectively improve the quasi-symmetry in the volume.
It is also notable how different the magnetic fields from the NAE and BoozerExact optimizations are: the Poincar\'e plots and rotational transform profiles drastically change.

The initial and final coils are shown in Figure \ref{fig:nae_coils}. To our surprise, the optimized coils found in this work visually overlap with the ones computed in Ref.~\onlinecite{giuliani_wechsung_stadler_cerfon_landreman_2022} which correspond to the precise QA configuration from Ref.~\onlinecite{PhysRevLett.128.035001}. 
In table \ref{tab:currents}, the coil currents found in this work and those in Ref.~\onlinecite{giuliani_wechsung_stadler_cerfon_landreman_2022} are provided, where the current in the first coil is normalized to 1.
We observe that the currents in the two designs are remarkably close to one another, even though the optimization algorithm starts with completely different initial coils.
\begin{table}
    \centering
    \begin{tabular}{c|c|c|c|c}
         &  Coil 1 & Coil 2 & Coil 3 & Coil 4\\
         \hline
         this work & 1.0000 &1.06831 & 1.4101 & 2.2341\\
         Ref.~\onlinecite{giuliani_wechsung_stadler_cerfon_landreman_2022}  & 1.0000 & 1.06370 & 1.4037 & 2.2508\\
    \end{tabular}
    \caption{Coil currents in the final designs of Ref.~\onlinecite{giuliani_wechsung_stadler_cerfon_landreman_2022} and this work, normalized so that the first coil's current is 1. 
    We observe that they are remarkably close to one another (within 1\%) even though the optimization algorithms start with completely different initial coils. 
    }
    \label{tab:currents}
\end{table}
Finding similar coils and magnetic field as in Ref.~\onlinecite{giuliani_wechsung_stadler_cerfon_landreman_2022} from scratch indicates that the surface optimization algorithm presented here is not restricted to stay in a neighborhood of the initial coil set and that the problem we solve here might suffer less from multiple minima.

\section{Conclusions and Discussion} \label{sec:discussion}
In this work, we have presented a general algorithm to robustly search for magnetic configurations with nested flux surfaces and precise QA, and which can be realized by coils.
This algorithm is a robust extension of our previous work in Ref.~\onlinecite{giuliani_wechsung_stadler_cerfon_landreman_2022} in the following sense: in Ref.~\onlinecite{giuliani_wechsung_stadler_cerfon_landreman_2022}, we only considered improvements of already optimized stellarator magnetic fields that had nested flux surfaces for a very large fraction of the total volume; in contrast, the extended algorithm we describe here performs well even when we use unoptimized magnetic fields as initial guesses for our optimizations.
The proposed approach first generates an initial coil set from either a FOCUS-like\cite{zhu2018designing, wechsung2022precise} or NAE optimization\cite{giuliani2022single}.
Next, we use the least squares surfaces formulation presented in this work to perform a physics optimization and reach an area of coil parameter space with nested flux surfaces.
This is done by fitting nested surfaces, even in regimes where flux surfaces do not exist. This is facilitated by the addition of a surface area regularization to avoid self-intersections.
After this optimization, which behaves robustly even in the presence of island or chaos, the stellarator design is improved using the local optimization algorithm described in Ref.~\onlinecite{giuliani_wechsung_stadler_cerfon_landreman_2022}.
In sections \ref{sec:coilopt-1} and \ref{sec:coilopt-2}, we show how to heal both localized islands and widespread chaos.
In section \ref{sec:coilopt-3}, we illustrate how to handle cold starts and show that our algorithm can lead to optimized magnetic configurations that are drastically different from the initial one.
Despite a completely different optimization algorithm and starting from flat coils and zero currents, we arrive at a coil set that is visually indistinguishable from the one obtained using the local algorithm in Ref.~\onlinecite{giuliani_wechsung_stadler_cerfon_landreman_2022}.
This hints that designing coils for QA on a volume is burdened by fewer local minima than targeting QA only on axis\cite{giuliani2022single}, or in a FOCUS-like optimization\cite{zhu2018designing, wechsung2022precise}.  This may also be due to the relative restrictive maximal coil length value $L_{\text{max}}$ we have chosen; a larger value might result in more local minima.  We plan on running more experiments to study this.  
In Table \ref{tab:geoprop}, we provide a comparison of the geometric properties of the stellarator designs presented in this work along with the geometric properties of their coils.
Consider first the island and chaos healing coil sets where we optimized up to surfaces with aspect ratio 6 and 4, respectively.
When including surfaces with lower aspect ratio in the optimization, the coil-to-surface separation decreases, and coil complexity (maximum curvature, mean-squared curvature) reduces.
At the expense of less precise quasi-symmetry, we obtain a design with lower coil complexity and nested flux surfaces with lower aspect ratio.
In all examples, the coil-coil separation inequality constraint was not active due to the chosen value of $L_{\max}=\SI{18}{\meter}$.  For longer coils, one can expect it to become active\cite{giuliani_wechsung_stadler_cerfon_landreman_2022}.

\begin{table*}\centering
    \begin{tabular}{r|p{1.25cm} p{3.25cm} p{3.25cm} p{3.25cm} p{1.9cm} p{1.9cm}}
        Coil set    & Aspect ratio & Coil lengths   & Maximum\newline curvatures & Mean-squared \newline curvatures & Coil-coil\newline separation & Coil-surf.\newline separation \\
        \hline
        Island healing & 5.99 & $4.08,5.37,4.37,4.17$ & $4.21,3.96,4.12,3.97$  &  $5.00,4.08,5.00,5.00$ &  $0.108$ & $0.254$ \\
        Chaos healing  & 4.09 & $3.93,5.05,4.85,4.17$ & $3.76,3.37,3.61,3.64$  &  $4.66,3.12,3.80,5.00$ &  $0.126$ & $0.195$ \\
        Cold start     & 5.99 & $5.37,3.94,4.10,4.60$ & $4.28,3.64,4.17,4.76$  &  $4.27,5.00,5.00,5.00$ &  $0.108$ & $0.240$  \\
    \end{tabular}
    \caption{Comparison of geometric properties of the stellarators and their coils. Coil-to-surface distance is computed with respect to the outermost surface on which quasi-symmetry is optimized.}\label{tab:geoprop}
\end{table*}

\begin{figure}
    \centering
    \includegraphics[]{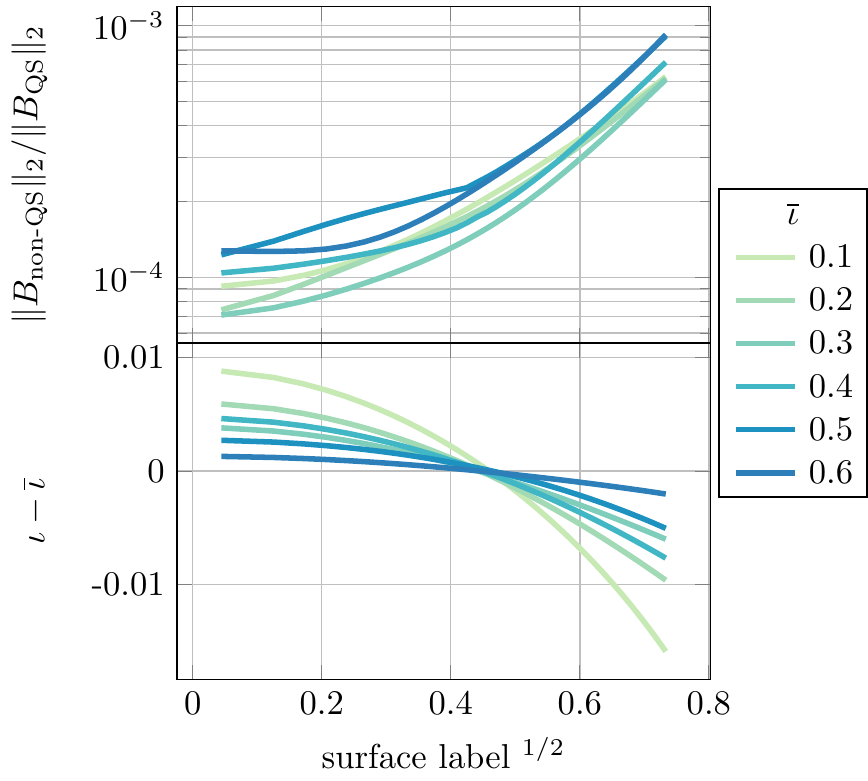}
    \caption{Quality of quasi-symmetry for various average rotational transforms $\overline\iota$. Shown of the top are the non-QA ratios.
    We are able to find stellarators with comparable quasi-symmetry for a wide range of average rotational transforms $\overline{\iota} = 0.1, 0.2, \hdots, 0.6$. Shown on the bottom is the deviation from the average rotational transforms.
    All runs use the shortest total allowable coil lengths $L_{\max} = \SI{18}{\meter}$ from the study in
    Ref.~\onlinecite{giuliani_wechsung_stadler_cerfon_landreman_2022}. For longer coils, one can expect improved quasi-symmetry, but this question is not explored in this article.
    }
    \label{fig:scan}
\end{figure}

This work opens the door to a more wider search for stellarator coils that produce nested flux surfaces with precise quasi-symmetry.  As a preliminary study, we use the same setup as in the cold start example (section \ref{sec:coilopt-3}), except that we consider various target average rotational transforms in the plasma volume, namely  $\overline{\iota}=0.1, 0.2, \hdots, 0.6$.  
Using the same initial coil set from a NAE optimization with $\iota=0.42$ on axis, our robust coil optimization procedure successfully finds coil sets with comparable quasi-symmetry for a wide range of rotational transform values (Figure \ref{fig:scan}), though larger rotational transforms corresponded to slightly worse QA.
Moreover, for the range of rotational transforms considered, we observe that magnetic shear generally decreases with increasing $\overline{\iota}$.
The scan of $\overline{\iota}$ completed here used a fixed length $L_{\max}=\SI{18}{\meter}$ for the sum of the 4 independent coils.
It is known that the total coil length is a strong regularizer for the outer optimization problem, and improved quasi-symmetry can be obtained by increasing $L_{\max}$, as shown in Ref.~\onlinecite{giuliani_wechsung_stadler_cerfon_landreman_2022}.
In this preliminary investigation, we always used the same initial coil set obtained from a single NAE optimization to illustrate our method's robustness and the ability to find coils for a variety of rotational transforms.
Alternatively, one might recompute the initial coil set from a NAE optimization for each $\overline{\iota}$ used in the parameter scan.

Using the tools presented in this work, we plan on completing a more detailed scan of target physics and coil design values: average rotational transform $\overline{\iota}$, $n_{\text{fp}}$, number of coils, maximum coil length $L_{\text{max}}$, aspect ratio, etc.
There are no fundamental restrictions that prevent extending this approach to other flavors of quasi-symmetry such as quasihelical or quasipoloidal symmetry.
We focused here on vacuum-field equilibria as they are an important aspect of stellarator optimization, and can be used to initialize other optimization studies without the curl-free assumption.
An important extension of this work is to apply these algorithms to direct coil optimization for finite $\beta$ magnetic equilibria.  Doing this requires the ability to compute the total magnetic field, along with its first and second spatial derivatives at a given point in space. 
Although not investigated here, it would also be interesting to develop an adaptive algorithm that modifies the number of collocation points and surface parameters on surfaces depending on the complexity of the surface.

\section*{Code availability and optimized configurations}
Many of the tools described here are available in SIMSOPT\cite{simsopt}, which is a suite of stellarator design, optimization, and analysis utilities:

\url{https://github.com/hiddenSymmetries/simsopt}

\section*{Acknowledgements}
The authors would like to thank the SIMSOPT development team. This work was
supported by a grant from the Simons Foundation (560651). AG is partially supported by
an NSERC (Natural Sciences and Engineering Research Council of Canada) postdoctoral
fellowship. In addition, AC and FW are supported by the United States National Science
Foundation under grant No.\ PHY-1820852, and AC is supported by the United States
Department of Energy, Office of Fusion Energy Sciences, under grant No.\ DE-FG02-
86ER53223. This work was supported in part through the NYU IT High Performance Computing resources, services, and staff expertise.
The authors also gratefully acknowledge helpful discussions with Elizabeth Paul.

\FloatBarrier

\bibliography{main}

\providecommand{\noopsort}[1]{}\providecommand{\singleletter}[1]{#1}%
\begin{thebibliography}{22}%
\makeatletter
\providecommand \@ifxundefined [1]{%
 \@ifx{#1\undefined}
}%
\providecommand \@ifnum [1]{%
 \ifnum #1\expandafter \@firstoftwo
 \else \expandafter \@secondoftwo
 \fi
}%
\providecommand \@ifx [1]{%
 \ifx #1\expandafter \@firstoftwo
 \else \expandafter \@secondoftwo
 \fi
}%
\providecommand \natexlab [1]{#1}%
\providecommand \enquote  [1]{``#1''}%
\providecommand \bibnamefont  [1]{#1}%
\providecommand \bibfnamefont [1]{#1}%
\providecommand \citenamefont [1]{#1}%
\providecommand \href@noop [0]{\@secondoftwo}%
\providecommand \href [0]{\begingroup \@sanitize@url \@href}%
\providecommand \@href[1]{\@@startlink{#1}\@@href}%
\providecommand \@@href[1]{\endgroup#1\@@endlink}%
\providecommand \@sanitize@url [0]{\catcode `\\12\catcode `\$12\catcode
  `\&12\catcode `\#12\catcode `\^12\catcode `\_12\catcode `\%12\relax}%
\providecommand \@@startlink[1]{}%
\providecommand \@@endlink[0]{}%
\providecommand \url  [0]{\begingroup\@sanitize@url \@url }%
\providecommand \@url [1]{\endgroup\@href {#1}{\urlprefix }}%
\providecommand \urlprefix  [0]{URL }%
\providecommand \Eprint [0]{\href }%
\providecommand \doibase [0]{http://dx.doi.org/}%
\providecommand \selectlanguage [0]{\@gobble}%
\providecommand \bibinfo  [0]{\@secondoftwo}%
\providecommand \bibfield  [0]{\@secondoftwo}%
\providecommand \translation [1]{[#1]}%
\providecommand \BibitemOpen [0]{}%
\providecommand \bibitemStop [0]{}%
\providecommand \bibitemNoStop [0]{.\EOS\space}%
\providecommand \EOS [0]{\spacefactor3000\relax}%
\providecommand \BibitemShut  [1]{\csname bibitem#1\endcsname}%
\let\auto@bib@innerbib\@empty
\bibitem [{\citenamefont {Giuliani}\ \emph
  {et~al.}(2022{\natexlab{a}})\citenamefont {Giuliani}, \citenamefont
  {Wechsung}, \citenamefont {Stadler}, \citenamefont {Cerfon},\ and\
  \citenamefont {Landreman}}]{giuliani_wechsung_stadler_cerfon_landreman_2022}%
  \BibitemOpen
  \bibfield  {author} {\bibinfo {author} {\bibfnamefont {A.}~\bibnamefont
  {Giuliani}}, \bibinfo {author} {\bibfnamefont {F.}~\bibnamefont {Wechsung}},
  \bibinfo {author} {\bibfnamefont {G.}~\bibnamefont {Stadler}}, \bibinfo
  {author} {\bibfnamefont {A.}~\bibnamefont {Cerfon}}, \ and\ \bibinfo {author}
  {\bibfnamefont {M.}~\bibnamefont {Landreman}},\ }\bibfield  {title} {\enquote
  {\bibinfo {title} {Direct computation of magnetic surfaces in {B}oozer
  coordinates and coil optimization for quasisymmetry},}\ }\href {\doibase
  10.1017/S0022377822000563} {\bibfield  {journal} {\bibinfo  {journal}
  {Journal of Plasma Physics}\ }\textbf {\bibinfo {volume} {88}},\ \bibinfo
  {pages} {905880401} (\bibinfo {year} {2022}{\natexlab{a}})}\BibitemShut
  {NoStop}%
\bibitem [{\citenamefont {Helander}(2014)}]{helander_review}%
  \BibitemOpen
  \bibfield  {author} {\bibinfo {author} {\bibfnamefont {P.}~\bibnamefont
  {Helander}},\ }\bibfield  {title} {\enquote {\bibinfo {title} {Theory of
  plasma confinement in non-axisymmetric magnetic fields},}\ }\href@noop {}
  {\bibfield  {journal} {\bibinfo  {journal} {Reports on Progress in Physics}\
  }\textbf {\bibinfo {volume} {77}},\ \bibinfo {pages} {087001} (\bibinfo
  {year} {2014})}\BibitemShut {NoStop}%
\bibitem [{\citenamefont {Landreman}\ and\ \citenamefont
  {Paul}(2022)}]{PhysRevLett.128.035001}%
  \BibitemOpen
  \bibfield  {author} {\bibinfo {author} {\bibfnamefont {M.}~\bibnamefont
  {Landreman}}\ and\ \bibinfo {author} {\bibfnamefont {E.}~\bibnamefont
  {Paul}},\ }\bibfield  {title} {\enquote {\bibinfo {title} {Magnetic fields
  with precise quasisymmetry for plasma confinement},}\ }\href {\doibase
  10.1103/PhysRevLett.128.035001} {\bibfield  {journal} {\bibinfo  {journal}
  {Phys. Rev. Lett.}\ }\textbf {\bibinfo {volume} {128}},\ \bibinfo {pages}
  {035001} (\bibinfo {year} {2022})}\BibitemShut {NoStop}%
\bibitem [{\citenamefont {Landreman}\ \emph {et~al.}(2021)\citenamefont
  {Landreman}, \citenamefont {Medasani}, \citenamefont {Wechsung},
  \citenamefont {Giuliani}, \citenamefont {Jorge},\ and\ \citenamefont
  {Zhu}}]{simsopt}%
  \BibitemOpen
  \bibfield  {author} {\bibinfo {author} {\bibfnamefont {M.}~\bibnamefont
  {Landreman}}, \bibinfo {author} {\bibfnamefont {B.}~\bibnamefont {Medasani}},
  \bibinfo {author} {\bibfnamefont {F.}~\bibnamefont {Wechsung}}, \bibinfo
  {author} {\bibfnamefont {A.}~\bibnamefont {Giuliani}}, \bibinfo {author}
  {\bibfnamefont {R.}~\bibnamefont {Jorge}}, \ and\ \bibinfo {author}
  {\bibfnamefont {C.}~\bibnamefont {Zhu}},\ }\bibfield  {title} {\enquote
  {\bibinfo {title} {{SIMSOPT:} a flexible framework for stellarator
  optimization},}\ }\href@noop {} {\bibfield  {journal} {\bibinfo  {journal}
  {Journal of Open Source Software}\ }\textbf {\bibinfo {volume} {6}},\
  \bibinfo {pages} {3525} (\bibinfo {year} {2021})}\BibitemShut {NoStop}%
\bibitem [{\citenamefont {Hudson}\ \emph {et~al.}(2002)\citenamefont {Hudson},
  \citenamefont {Reiman}, \citenamefont {Strickler}, \citenamefont {Brooks},
  \citenamefont {Monticello},\ and\ \citenamefont {Hirshman}}]{Hudson_2002}%
  \BibitemOpen
  \bibfield  {author} {\bibinfo {author} {\bibfnamefont {S.~R.}\ \bibnamefont
  {Hudson}}, \bibinfo {author} {\bibfnamefont {A.}~\bibnamefont {Reiman}},
  \bibinfo {author} {\bibfnamefont {D.}~\bibnamefont {Strickler}}, \bibinfo
  {author} {\bibfnamefont {A.}~\bibnamefont {Brooks}}, \bibinfo {author}
  {\bibfnamefont {D.~A.}\ \bibnamefont {Monticello}}, \ and\ \bibinfo {author}
  {\bibfnamefont {S.~P.}\ \bibnamefont {Hirshman}},\ }\bibfield  {title}
  {\enquote {\bibinfo {title} {Free-boundary full-pressure island healing in
  stellarator equilibria: coil-healing},}\ }\href {\doibase
  10.1088/0741-3335/44/7/323} {\bibfield  {journal} {\bibinfo  {journal}
  {Plasma Physics and Controlled Fusion}\ }\textbf {\bibinfo {volume} {44}},\
  \bibinfo {pages} {1377--1382} (\bibinfo {year} {2002})}\BibitemShut {NoStop}%
\bibitem [{\citenamefont {Hudson}\ \emph {et~al.}(2003)\citenamefont {Hudson},
  \citenamefont {Monticello}, \citenamefont {Reiman}, \citenamefont
  {Strickler}, \citenamefont {Hirshman}, \citenamefont {Ku}, \citenamefont
  {Lazarus}, \citenamefont {Brooks}, \citenamefont {Zarnstorff}, \citenamefont
  {Boozer}, \citenamefont {Fu},\ and\ \citenamefont {Neilson}}]{Hudson_2003}%
  \BibitemOpen
  \bibfield  {author} {\bibinfo {author} {\bibfnamefont {S.}~\bibnamefont
  {Hudson}}, \bibinfo {author} {\bibfnamefont {D.}~\bibnamefont {Monticello}},
  \bibinfo {author} {\bibfnamefont {A.}~\bibnamefont {Reiman}}, \bibinfo
  {author} {\bibfnamefont {D.}~\bibnamefont {Strickler}}, \bibinfo {author}
  {\bibfnamefont {S.}~\bibnamefont {Hirshman}}, \bibinfo {author}
  {\bibfnamefont {L.-P.}\ \bibnamefont {Ku}}, \bibinfo {author} {\bibfnamefont
  {E.}~\bibnamefont {Lazarus}}, \bibinfo {author} {\bibfnamefont
  {A.}~\bibnamefont {Brooks}}, \bibinfo {author} {\bibfnamefont
  {M.}~\bibnamefont {Zarnstorff}}, \bibinfo {author} {\bibfnamefont
  {A.}~\bibnamefont {Boozer}}, \bibinfo {author} {\bibfnamefont {G.-Y.}\
  \bibnamefont {Fu}}, \ and\ \bibinfo {author} {\bibfnamefont {G.}~\bibnamefont
  {Neilson}},\ }\bibfield  {title} {\enquote {\bibinfo {title} {Constructing
  integrable high-pressure full-current free-boundary stellarator
  magnetohydrodynamic equilibrium solutions},}\ }\href {\doibase
  10.1088/0029-5515/43/10/004} {\bibfield  {journal} {\bibinfo  {journal}
  {Nuclear Fusion}\ }\textbf {\bibinfo {volume} {43}},\ \bibinfo {pages}
  {1040--1046} (\bibinfo {year} {2003})}\BibitemShut {NoStop}%
\bibitem [{\citenamefont {Zhu}\ \emph {et~al.}(2019)\citenamefont {Zhu},
  \citenamefont {Gates}, \citenamefont {Hudson}, \citenamefont {Liu},
  \citenamefont {Xu}, \citenamefont {Shimizu},\ and\ \citenamefont
  {Okamura}}]{Zhu_2019}%
  \BibitemOpen
  \bibfield  {author} {\bibinfo {author} {\bibfnamefont {C.}~\bibnamefont
  {Zhu}}, \bibinfo {author} {\bibfnamefont {D.~A.}\ \bibnamefont {Gates}},
  \bibinfo {author} {\bibfnamefont {S.~R.}\ \bibnamefont {Hudson}}, \bibinfo
  {author} {\bibfnamefont {H.}~\bibnamefont {Liu}}, \bibinfo {author}
  {\bibfnamefont {Y.}~\bibnamefont {Xu}}, \bibinfo {author} {\bibfnamefont
  {A.}~\bibnamefont {Shimizu}}, \ and\ \bibinfo {author} {\bibfnamefont
  {S.}~\bibnamefont {Okamura}},\ }\bibfield  {title} {\enquote {\bibinfo
  {title} {Identification of important error fields in stellarators using the
  {H}essian matrix method},}\ }\href {\doibase 10.1088/1741-4326/ab3a7c}
  {\bibfield  {journal} {\bibinfo  {journal} {Nuclear Fusion}\ }\textbf
  {\bibinfo {volume} {59}},\ \bibinfo {pages} {126007} (\bibinfo {year}
  {2019})}\BibitemShut {NoStop}%
\bibitem [{\citenamefont {Landreman}, \citenamefont {Medasani},\ and\
  \citenamefont {Zhu}(2021)}]{landreman2021stellarator}%
  \BibitemOpen
  \bibfield  {author} {\bibinfo {author} {\bibfnamefont {M.}~\bibnamefont
  {Landreman}}, \bibinfo {author} {\bibfnamefont {B.}~\bibnamefont {Medasani}},
  \ and\ \bibinfo {author} {\bibfnamefont {C.}~\bibnamefont {Zhu}},\ }\bibfield
   {title} {\enquote {\bibinfo {title} {Stellarator optimization for good
  magnetic surfaces at the same time as quasisymmetry},}\ }\href@noop {}
  {\bibfield  {journal} {\bibinfo  {journal} {Physics of Plasmas}\ }\textbf
  {\bibinfo {volume} {28}},\ \bibinfo {pages} {092505} (\bibinfo {year}
  {2021})}\BibitemShut {NoStop}%
\bibitem [{\citenamefont {Baillod}\ \emph {et~al.}(2022)\citenamefont
  {Baillod}, \citenamefont {Loizu}, \citenamefont {Graves},\ and\ \citenamefont
  {Landreman}}]{baillod2022stellarator}%
  \BibitemOpen
  \bibfield  {author} {\bibinfo {author} {\bibfnamefont {A.}~\bibnamefont
  {Baillod}}, \bibinfo {author} {\bibfnamefont {J.}~\bibnamefont {Loizu}},
  \bibinfo {author} {\bibfnamefont {J.}~\bibnamefont {Graves}}, \ and\ \bibinfo
  {author} {\bibfnamefont {M.}~\bibnamefont {Landreman}},\ }\bibfield  {title}
  {\enquote {\bibinfo {title} {Stellarator optimization for nested magnetic
  surfaces at finite $\beta$ and toroidal current},}\ }\href@noop {} {\bibfield
   {journal} {\bibinfo  {journal} {Physics of Plasmas}\ }\textbf {\bibinfo
  {volume} {29}},\ \bibinfo {pages} {042505} (\bibinfo {year}
  {2022})}\BibitemShut {NoStop}%
\bibitem [{\citenamefont {Cary}(1982)}]{cary1982vacuum}%
  \BibitemOpen
  \bibfield  {author} {\bibinfo {author} {\bibfnamefont {J.~R.}\ \bibnamefont
  {Cary}},\ }\bibfield  {title} {\enquote {\bibinfo {title} {Vacuum magnetic
  fields with dense flux surfaces},}\ }\href@noop {} {\bibfield  {journal}
  {\bibinfo  {journal} {Physical Review Letters}\ }\textbf {\bibinfo {volume}
  {49}},\ \bibinfo {pages} {276} (\bibinfo {year} {1982})}\BibitemShut
  {NoStop}%
\bibitem [{\citenamefont {Dommaschk}(1982)}]{dommaschk1982finite}%
  \BibitemOpen
  \bibfield  {author} {\bibinfo {author} {\bibfnamefont {W.}~\bibnamefont
  {Dommaschk}},\ }\bibfield  {title} {\enquote {\bibinfo {title} {Finite field
  harmonics for stellarators with improved aspect ratio},}\ }\href@noop {}
  {\bibfield  {journal} {\bibinfo  {journal} {Zeitschrift f{\"u}r
  Naturforschung A}\ }\textbf {\bibinfo {volume} {37}},\ \bibinfo {pages}
  {866--875} (\bibinfo {year} {1982})}\BibitemShut {NoStop}%
\bibitem [{\citenamefont {Cary}(1984)}]{cary1984construction}%
  \BibitemOpen
  \bibfield  {author} {\bibinfo {author} {\bibfnamefont {J.~R.}\ \bibnamefont
  {Cary}},\ }\bibfield  {title} {\enquote {\bibinfo {title} {Construction of
  three-dimensional vacuum magnetic fields with dense nested flux surfaces},}\
  }\href@noop {} {\bibfield  {journal} {\bibinfo  {journal} {The Physics of
  fluids}\ }\textbf {\bibinfo {volume} {27}},\ \bibinfo {pages} {119--128}
  (\bibinfo {year} {1984})}\BibitemShut {NoStop}%
\bibitem [{\citenamefont {Hanson}\ and\ \citenamefont
  {Cary}(1984)}]{Hanson_1984_stochasticity}%
  \BibitemOpen
  \bibfield  {author} {\bibinfo {author} {\bibfnamefont {J.~D.}\ \bibnamefont
  {Hanson}}\ and\ \bibinfo {author} {\bibfnamefont {J.~R.}\ \bibnamefont
  {Cary}},\ }\bibfield  {title} {\enquote {\bibinfo {title} {Elimination of
  stochasticity in stellarators},}\ }\href {\doibase 10.1063/1.864692}
  {\bibfield  {journal} {\bibinfo  {journal} {The Physics of Fluids}\ }\textbf
  {\bibinfo {volume} {27}},\ \bibinfo {pages} {767--769} (\bibinfo {year}
  {1984})},\ \Eprint
  {http://arxiv.org/abs/https://aip.scitation.org/doi/pdf/10.1063/1.864692}
  {https://aip.scitation.org/doi/pdf/10.1063/1.864692} \BibitemShut {NoStop}%
\bibitem [{\citenamefont {Cary}\ and\ \citenamefont
  {Hanson}(1986)}]{Cary1986_stochasticity}%
  \BibitemOpen
  \bibfield  {author} {\bibinfo {author} {\bibfnamefont {J.~R.}\ \bibnamefont
  {Cary}}\ and\ \bibinfo {author} {\bibfnamefont {J.~D.}\ \bibnamefont
  {Hanson}},\ }\bibfield  {title} {\enquote {\bibinfo {title} {Stochasticity
  reduction},}\ }\href {\doibase 10.1063/1.865539} {\bibfield  {journal}
  {\bibinfo  {journal} {The Physics of Fluids}\ }\textbf {\bibinfo {volume}
  {29}},\ \bibinfo {pages} {2464--2473} (\bibinfo {year} {1986})},\ \Eprint
  {http://arxiv.org/abs/https://aip.scitation.org/doi/pdf/10.1063/1.865539}
  {https://aip.scitation.org/doi/pdf/10.1063/1.865539} \BibitemShut {NoStop}%
\bibitem [{\citenamefont {Lee}\ \emph {et~al.}(2022)\citenamefont {Lee},
  \citenamefont {Paul}, \citenamefont {Stadler},\ and\ \citenamefont
  {Landreman}}]{ejpauliota}%
  \BibitemOpen
  \bibfield  {author} {\bibinfo {author} {\bibfnamefont {B.~F.}\ \bibnamefont
  {Lee}}, \bibinfo {author} {\bibfnamefont {E.~J.}\ \bibnamefont {Paul}},
  \bibinfo {author} {\bibfnamefont {G.}~\bibnamefont {Stadler}}, \ and\
  \bibinfo {author} {\bibfnamefont {M.}~\bibnamefont {Landreman}},\ }\href
  {\doibase 10.48550/ARXIV.2208.01096} {\enquote {\bibinfo {title} {Stellarator
  coil optimization supporting multiple magnetic configurations},}\ } (\bibinfo
  {year} {2022})\BibitemShut {NoStop}%
\bibitem [{\citenamefont {Dudt}\ \emph {et~al.}(2022)\citenamefont {Dudt},
  \citenamefont {Conlin}, \citenamefont {Panici},\ and\ \citenamefont
  {Kolemen}}]{desc}%
  \BibitemOpen
  \bibfield  {author} {\bibinfo {author} {\bibfnamefont {D.}~\bibnamefont
  {Dudt}}, \bibinfo {author} {\bibfnamefont {R.}~\bibnamefont {Conlin}},
  \bibinfo {author} {\bibfnamefont {D.}~\bibnamefont {Panici}}, \ and\ \bibinfo
  {author} {\bibfnamefont {E.}~\bibnamefont {Kolemen}},\ }\href {\doibase
  10.48550/ARXIV.2204.00078} {\enquote {\bibinfo {title} {The {DESC}
  stellarator code suite part iii: Quasi-symmetry optimization},}\ } (\bibinfo
  {year} {2022})\BibitemShut {NoStop}%
\bibitem [{\citenamefont {Boozer}(2019)}]{boozer2019curl}%
  \BibitemOpen
  \bibfield  {author} {\bibinfo {author} {\bibfnamefont {A.~H.}\ \bibnamefont
  {Boozer}},\ }\bibfield  {title} {\enquote {\bibinfo {title} {Curl-free
  magnetic fields for stellarator optimization},}\ }\href@noop {} {\bibfield
  {journal} {\bibinfo  {journal} {Physics of Plasmas}\ }\textbf {\bibinfo
  {volume} {26}},\ \bibinfo {pages} {102504} (\bibinfo {year}
  {2019})}\BibitemShut {NoStop}%
\bibitem [{\citenamefont {Dewar}, \citenamefont {Hudson},\ and\ \citenamefont
  {Price}(1994)}]{DEWAR199449}%
  \BibitemOpen
  \bibfield  {author} {\bibinfo {author} {\bibfnamefont {R.}~\bibnamefont
  {Dewar}}, \bibinfo {author} {\bibfnamefont {S.}~\bibnamefont {Hudson}}, \
  and\ \bibinfo {author} {\bibfnamefont {P.}~\bibnamefont {Price}},\ }\bibfield
   {title} {\enquote {\bibinfo {title} {Almost invariant manifolds for
  divergence-free fields},}\ }\href {\doibase
  https://doi.org/10.1016/0375-9601(94)00707-V} {\bibfield  {journal} {\bibinfo
   {journal} {Physics Letters A}\ }\textbf {\bibinfo {volume} {194}},\ \bibinfo
  {pages} {49--56} (\bibinfo {year} {1994})}\BibitemShut {NoStop}%
\bibitem [{\citenamefont {Giuliani}\ \emph
  {et~al.}(2022{\natexlab{b}})\citenamefont {Giuliani}, \citenamefont
  {Wechsung}, \citenamefont {Cerfon}, \citenamefont {Stadler},\ and\
  \citenamefont {Landreman}}]{giuliani2022single}%
  \BibitemOpen
  \bibfield  {author} {\bibinfo {author} {\bibfnamefont {A.}~\bibnamefont
  {Giuliani}}, \bibinfo {author} {\bibfnamefont {F.}~\bibnamefont {Wechsung}},
  \bibinfo {author} {\bibfnamefont {A.}~\bibnamefont {Cerfon}}, \bibinfo
  {author} {\bibfnamefont {G.}~\bibnamefont {Stadler}}, \ and\ \bibinfo
  {author} {\bibfnamefont {M.}~\bibnamefont {Landreman}},\ }\bibfield  {title}
  {\enquote {\bibinfo {title} {Single-stage gradient-based stellarator coil
  design: {O}ptimization for near-axis quasi-symmetry},}\ }\href@noop {}
  {\bibfield  {journal} {\bibinfo  {journal} {Journal of Computational
  Physics}\ }\textbf {\bibinfo {volume} {459}},\ \bibinfo {pages} {111147}
  (\bibinfo {year} {2022}{\natexlab{b}})}\BibitemShut {NoStop}%
\bibitem [{\citenamefont {Landreman}\ and\ \citenamefont
  {Boozer}(2016)}]{landreman2016efficient}%
  \BibitemOpen
  \bibfield  {author} {\bibinfo {author} {\bibfnamefont {M.}~\bibnamefont
  {Landreman}}\ and\ \bibinfo {author} {\bibfnamefont {A.~H.}\ \bibnamefont
  {Boozer}},\ }\bibfield  {title} {\enquote {\bibinfo {title} {Efficient
  magnetic fields for supporting toroidal plasmas},}\ }\href@noop {} {\bibfield
   {journal} {\bibinfo  {journal} {Physics of Plasmas}\ }\textbf {\bibinfo
  {volume} {23}},\ \bibinfo {pages} {032506} (\bibinfo {year}
  {2016})}\BibitemShut {NoStop}%
\bibitem [{\citenamefont {Zhu}\ \emph {et~al.}(2018)\citenamefont {Zhu},
  \citenamefont {Hudson}, \citenamefont {Song},\ and\ \citenamefont
  {Wan}}]{zhu2018designing}%
  \BibitemOpen
  \bibfield  {author} {\bibinfo {author} {\bibfnamefont {C.}~\bibnamefont
  {Zhu}}, \bibinfo {author} {\bibfnamefont {S.~R.}\ \bibnamefont {Hudson}},
  \bibinfo {author} {\bibfnamefont {Y.}~\bibnamefont {Song}}, \ and\ \bibinfo
  {author} {\bibfnamefont {Y.}~\bibnamefont {Wan}},\ }\bibfield  {title}
  {\enquote {\bibinfo {title} {Designing stellarator coils by a modified
  {N}ewton method using {FOCUS}},}\ }\href@noop {} {\bibfield  {journal}
  {\bibinfo  {journal} {Plasma Physics and Controlled Fusion}\ }\textbf
  {\bibinfo {volume} {60}},\ \bibinfo {pages} {065008} (\bibinfo {year}
  {2018})}\BibitemShut {NoStop}%
\bibitem [{\citenamefont {Wechsung}\ \emph {et~al.}(2022)\citenamefont
  {Wechsung}, \citenamefont {Landreman}, \citenamefont {Giuliani},
  \citenamefont {Cerfon},\ and\ \citenamefont {Stadler}}]{wechsung2022precise}%
  \BibitemOpen
  \bibfield  {author} {\bibinfo {author} {\bibfnamefont {F.}~\bibnamefont
  {Wechsung}}, \bibinfo {author} {\bibfnamefont {M.}~\bibnamefont {Landreman}},
  \bibinfo {author} {\bibfnamefont {A.}~\bibnamefont {Giuliani}}, \bibinfo
  {author} {\bibfnamefont {A.}~\bibnamefont {Cerfon}}, \ and\ \bibinfo {author}
  {\bibfnamefont {G.}~\bibnamefont {Stadler}},\ }\bibfield  {title} {\enquote
  {\bibinfo {title} {Precise stellarator quasi-symmetry can be achieved with
  electromagnetic coils},}\ }\href@noop {} {\bibfield  {journal} {\bibinfo
  {journal} {Proceedings of the National Academy of Sciences}\ }\textbf
  {\bibinfo {volume} {119}},\ \bibinfo {pages} {e2202084119} (\bibinfo {year}
  {2022})}\BibitemShut {NoStop}%
\end{thebibliography}%

\end{document}